\documentclass[12pt]{iopart}
\usepackage{graphicx}
\usepackage{iopams}
\usepackage{epsf}

\begin{document}

\newcommand{\rem}[1]{{\bf #1}}

















\hfill TU-791

\title{Cosmological Constraints on Neutrino Injection}

\author{Toru Kanzaki, Masahiro Kawasaki}
\address{Institute for Cosmic Ray Research,
 University of Tokyo, Kashiwa 277-8582, JAPAN}
\author{Kazunori Kohri}
\address{Physics Department, Lancaster University, Lancaster LA1 4YB,
 UK}
\author{Takeo Moroi}
\address{Department of Physics, Tohoku University,
 Sendai 980-8578, JAPAN}

\begin{abstract}

We derive general constraints on the relic abundances of a long-lived
particle which mainly decays into a neutrino (and something else) at
cosmological time scales.  Such an exotic particle may show up in
various particle-physics models based on physics beyond the standard
model.  The constraints are obtained from big-bang nucleosynthesis,
cosmic microwave background and diffuse neutrino and photon fluxes,
depending on the lifetime and the electromagnetic and hadronic branching
ratios.

\end{abstract}



\section{Introduction}
\label{sec:intro}

In modern cosmology, success of the big-bang nucleosynthesis (BBN) 
and existence of the cosmic microwave background (CMB) 
are important facts that support the standard big-bang model.
Prediction of the standard BBN scenario is in a reasonable 
agreement with the observations and 
the COBE observations~\cite{FIRAS} showed the perfect blackbody 
of CMB spectrum. 
On the other hand, in particle physics, if we consider physics 
beyond the standard model, there exist various new particles 
some of which have long lifetimes and decay during or after BBN. 
Examples of such long-lived particles include gravitino and moduli
predicted in the framework of supersymmetry (SUSY) and string theories.

BBN and CMB are useful probes to exotic particles predicted in physics
beyond standard model.  In fact, the prediction of BBN changes
significantly if there exists an exotic massive particle with long
lifetime.  (Hereafter, we call such a particle $X$.)  When the lifetime
of $X$ is longer than about 1 sec, the decay of $X$ may induce
electromagnetic and hadronic showers, which lead to photo- and
hadro-dissociation of ${\rm ^4He}$ and subsequent non-thermal production
of other light elements (${\rm D}$, ${\rm ^3He}$, ${\rm ^6Li}$, and
${\rm ^7Li}$).  Such processes may significantly change the prediction
of the standard BBN scenario and, consequently, resultant abundances of
light elements may conflict with observations.  Furthermore, the
electromagnetic energy injection causes distortion of the CMB spectrum.
Since the observation~\cite{FIRAS} shows that this distortion is quite
small, we can constrain the abundance of $X$.  Finally, if the lifetime
is very long, the spectrum of neutrinos and photons produced by the
decay of $X$ are not thermalized and may be directly observed.

The effects of the long-lived particles on BBN were well studied for
radiative decay~\cite{BBNwX_OLD, KMrad,Holtmann:1998gd,Rad_recent} and
for hadronic decay~\cite{Dominguez:1987,
Reno:1987qw,HadronicDecay,Kohri:2001jx,KKM04,Kohri:2005wn,Jedamzik:2006xz}
and stringent constraints on the abundance and lifetime of $X$ were
obtained. However, when $X$ mainly decays into neutrinos, it is expected
that the constraints become much weaker because of weakness of
interactions between neutrinos and other standard model particles.  The
specific case where a sneutrino mainly decays into a gravitino and a
neutrino has been already discussed in
\cite{Kawasaki:1994bs,FenSuTak,Kanzaki:2006hm}.  Such a scenario is
realized when the sneutrino is the next lightest superparticle (NLSP)
while the gravitino is the lightest superparticle (LSP).\footnote
{For the case that the gravitino is the the NLSP and that sneutrino is
  the LSP, see also \cite{Kawasaki:1994bs}.}
In such a case, interaction of $X$ (i.e., sneutrino) is well-known,
and it is found that BBN provides the most stringent constraint.

However, there are other possibilities of having long-lived massive
particles which dominantly decay into  neutrinos.  For example, if the
LSP is axino ($\tilde{a}$) and the NLSP is sneutrino, dominant 
decay process of the
sneutrino ($=X$) is $\tilde{\nu}\to\nu+\tilde{a}$.  
Decay rate of this process
depends on the properties of axion supermultiplet.  Thus, in general,
properties of $X$ (i.e., lifetime, hadronic branching ratio, and so
on) is model-dependent.  Consequently, the most stringent bound may not
be from BBN.  For example, high energy neutrinos emitted in the $X$
particle decay was considered in \cite{Gondolo:1991rn} where the upper
bounds on the $X$ abundance were obtained from nucleon-decay detectors
and Fly's Eye air shower array.

In this paper, we derive general cosmological constraints on scenarios
in which there exists a long-lived massive particle which dominantly
decays into a neutrino (and something else).  We treat the lifetime and
hadronic branching ratio of the long-lived particle as free
parameters.  In this case, in fact, the constraint from the main decay
mode is quite weak, and other subdominant decay channel which contain
electronic and hadronic particles may be
important~\cite{Kanzaki:2006hm}.  In our analysis, we take into
account both of these decay channels and discuss various cosmological
constraints.

Organization of this paper is as followis.  In Section \ref{sec:bbn},
we discuss BBN constraints.  In Section \ref{sec:cmb}, constraints
from CMB is considered.  Then in Section \ref{sec:diffuse}, we study
constraints from diffuse neutrino and photon.  Section
\ref{sec:conclusion} is devoted for conclusion.

\section{BBN Constraints}
\label{sec:bbn}

Before going into the main subject of this section, which is the
constraints from the BBN, let us first summarize the properties of the
long-lived heavy particle $X$ which mainly decays into a neutrino (and
some other weakly interacting particle).  In our study, we assume that
$X$ mainly decays as
\begin{eqnarray}
 X \rightarrow \nu + Y,
\end{eqnarray}
where $Y$ is an invisible particle which is very weakly interacting so
that it does not cause any subsequent scattering with background
particles.  One of the well-motivated examples is the case where the
sneutrino is the NLSP while gravitino (or axino) is the LSP.  (Then,
$X$ is the sneutrino and $Y$ is the gravitino or axino.)  In addition,
for concreteness, we assume that the final-state neutrino is electron
neutrino.  (We note here that we have checked that the constraints on
the properties of $X$ are not sensitive to the flavor of the
final-state neutrino.)

Even though the dominant decay mode is the two-body process, one
should keep in mind that decay chennels with three- and/or four-body
final state should also exist since the neutrino as well as $X$ and/or
$Y$ couple to $Z$- and $W$-bosons.  The emitted (real or virtual) weak
bosons subsequently decay into quarks and leptons.  With this type of
three- and/or four-body decay processes, energetic quarks and charged
leptons are produced.

If the decays of $X$ occur during or after BBN, the standard-model
particles emitted in the decay can affect the abundances of primordial
light elements.  First the high energy neutrinos emitted in the main
decay mode (two-body decay) scatter off the background leptons and
produce charged leptons ($e^{\pm}$, $\mu^{\pm}$) and charged
pions. The former induce electromagnetic showers which destroy light
elements, and the latter change the $n$-$p$ ratio through nucleon and
pion interactions.  Second $X$ decays into electromagnetic and
hadronic particles via the three- and/or four-body decay modes with
small branching ratio $B_X$.  Such processes directly induce
electromagnetic and hadronic showers and change the abundances of
light elements~\cite{KKM04}.

\subsection{Two-body Decay}

First, we discuss effects of the dominant decay process $X\to Y+\nu_e$.
We presume that $Y$ produced in the decay is a very weakly interacting
particle, and that it is irrelevant for BBN.  Neutrino, however, may
affect abundances of light elements.  The emitted energetic neutrinos
scatter off background leptons via weak interaction and several
kinds of particles may be pair-produced.

First, charged leptons may be produced via the following
processes:\footnote
{Neutrinos may also scatter off the background electron and positron.
  However, since the photo-dissociation processes become important
  when the cosmic temperature becomes much lower than $1\ {\rm MeV}$.
  At such temperature, number densities of electron and positron are
  exremely suppressed, and hence the scattering processes with
  electron and positron are irrelevant for the production of charged
  leptons. }
\begin{eqnarray}
 \nu_i+\bar{\nu}_{i,{\rm BG}}&\to&e^{-}+e^{+}
  \label{nunu2ee}\\
 \nu_i+\bar{\nu}_{i,{\rm BG}}&\to&\mu^{-}+\mu^{+}
  \\
 \nu_\mu+\bar{\nu}_{e,{\rm BG}}&\to&\mu^{-}+e^{+}
  \\
 \nu_e+\bar{\nu}_{\mu,{\rm BG}}&\to&e^{-}+\mu^{+}
  \label{nunu2emu}
\end{eqnarray}  
where $i=e,\mu,\tau$ is flavor index, and the subscript ``BG'' is for
background particles.  The muons emitted in the above processes quickly
decay into electrons and neutrinos.  Thus, the above processes produce
energetic electrons and positrons which cause electromagnetic cascade.
Energetic photons in the cascade induce photo-dissociation processes
of light elements.  Effects of these processes have been already
studied in~\cite{Kawasaki:1994bs}.

Another possible effect is due to the production of pion pairs.
High energy neutrinos scatter off the background neutrinos 
and electrons (positrons) and produce pions as
\begin{eqnarray}
  \nu_i + \bar{\nu}_{i,{\rm BG}}&\to&\pi^{-}+ \pi^{+},
  \\
  \nu_i + {e}^{\pm}_{i,{\rm BG}}&\to&\pi^{0}+ \pi^{\pm}.
\end{eqnarray}
The nucleus-pion interaction rate is $\sim 10^8 {\rm sec}^{-1}\times
(T/{\rm MeV})^3$ which is larger than the decay rate of the charged
pion $(\sim 4\times10^7 {\rm sec}^{-1})$ for $T\sim 1 {\rm MeV}$.
Therefore, the charged pions produced at $T\sim1 {\rm MeV}$ scatter
off the background nuclei and change protons (neutrons) into neutrons
(protons) via
\begin{eqnarray}
        \pi^{-}+p&\to&n+\pi^{0},~n+\gamma,
        \\
        \pi^{+}+n&\to&p+\pi^{0},~p+\gamma.
\end{eqnarray}  
Consequently, $n/p$ ratio is increased, resulting in more ${}^4{\rm
  He}$.  Notice that, because of very short lifetime, the neutral
pions decay before they scatter off the background nuclei and hence
they are harmless.

In order to estimate effects of the high-energy-neutrino induced
processes, we have numerically solved the Boltzmann equation describing
the time evolution of the high energy neutrino spectrum taking into
account of all the processes above.  Details of our calculation are
given in~\ref{app:Boltzmann}.

\subsection{Three and Four-body Decay}

Even though the branching ratio for three and four-body decay
processes are much smaller than 1, such decay processes are very
important since colored and charged particles are directly emitted
from these decay processes.  Energetic colored and charged particles
may significantly change the prediction of the standard BBN scenario.
Effects of these particles are classified into
three categories: photo-dissociations, hadro-dissociations and
$p\leftrightarrow n$ conversion.

\begin{figure}[t]
        \begin{center}
                \includegraphics[width=0.8\linewidth]{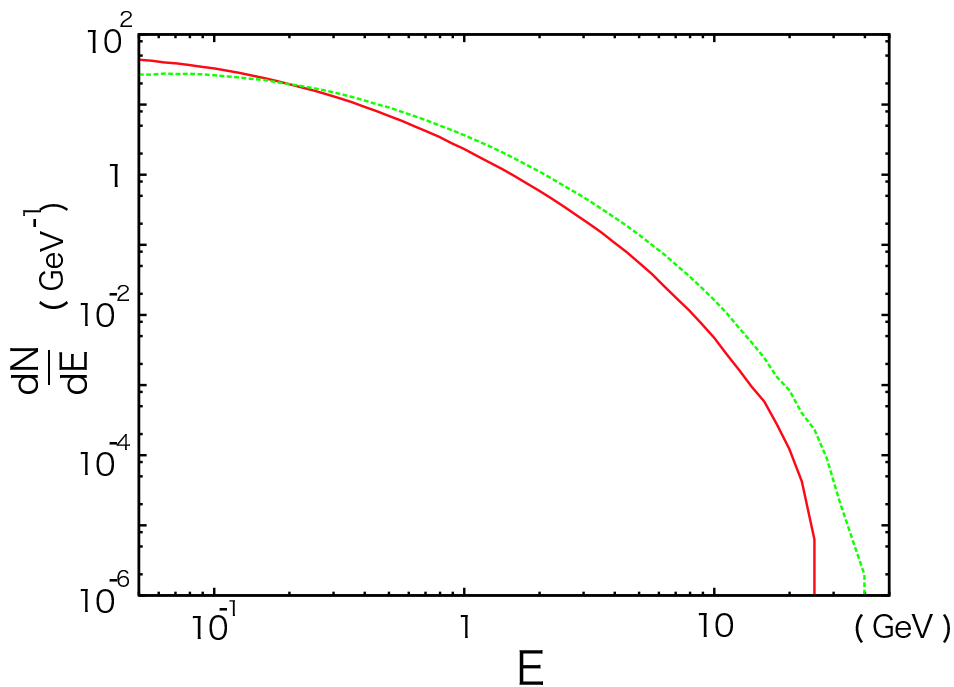}
                \caption{Spectrum of $e^+ + e^-$ (solid line) and 
                photon (dashed line) with $m_X=10^2{\rm GeV}$.}
        \label{fig:decay_m=1e2}
        \end{center}
\end{figure}

\begin{figure}[htbp]
        \begin{center}
                \includegraphics[width=0.8\linewidth]{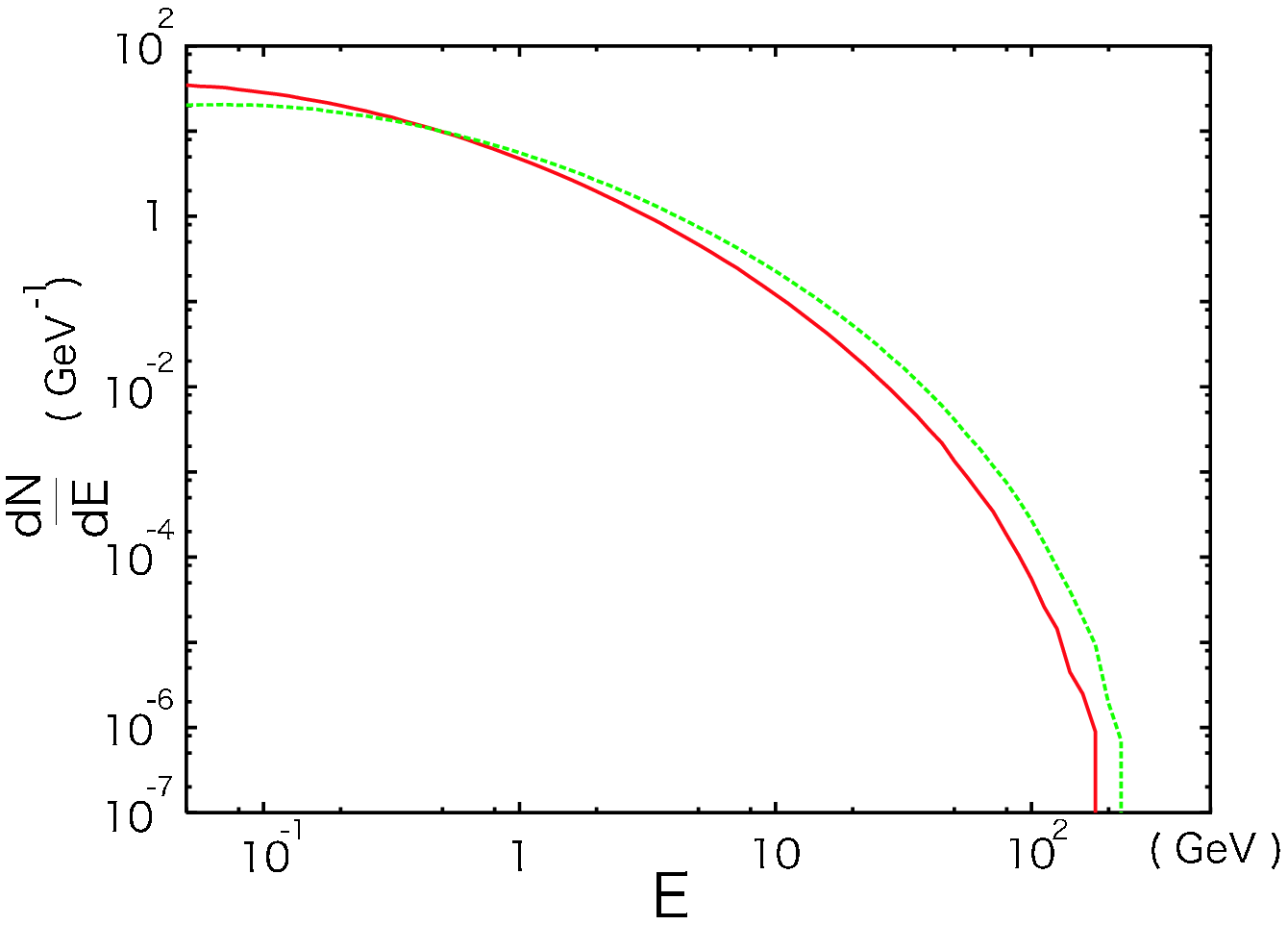}
                \caption{Same as Fig.\ref{fig:decay_m=1e2} except
                 $m_X=10^3{\rm GeV}$.}
        \label{fig:decay_m=1e3}
        \end{center}
\end{figure}

\begin{figure}[htbp]
        \begin{center}
                \includegraphics[width=0.8\linewidth]{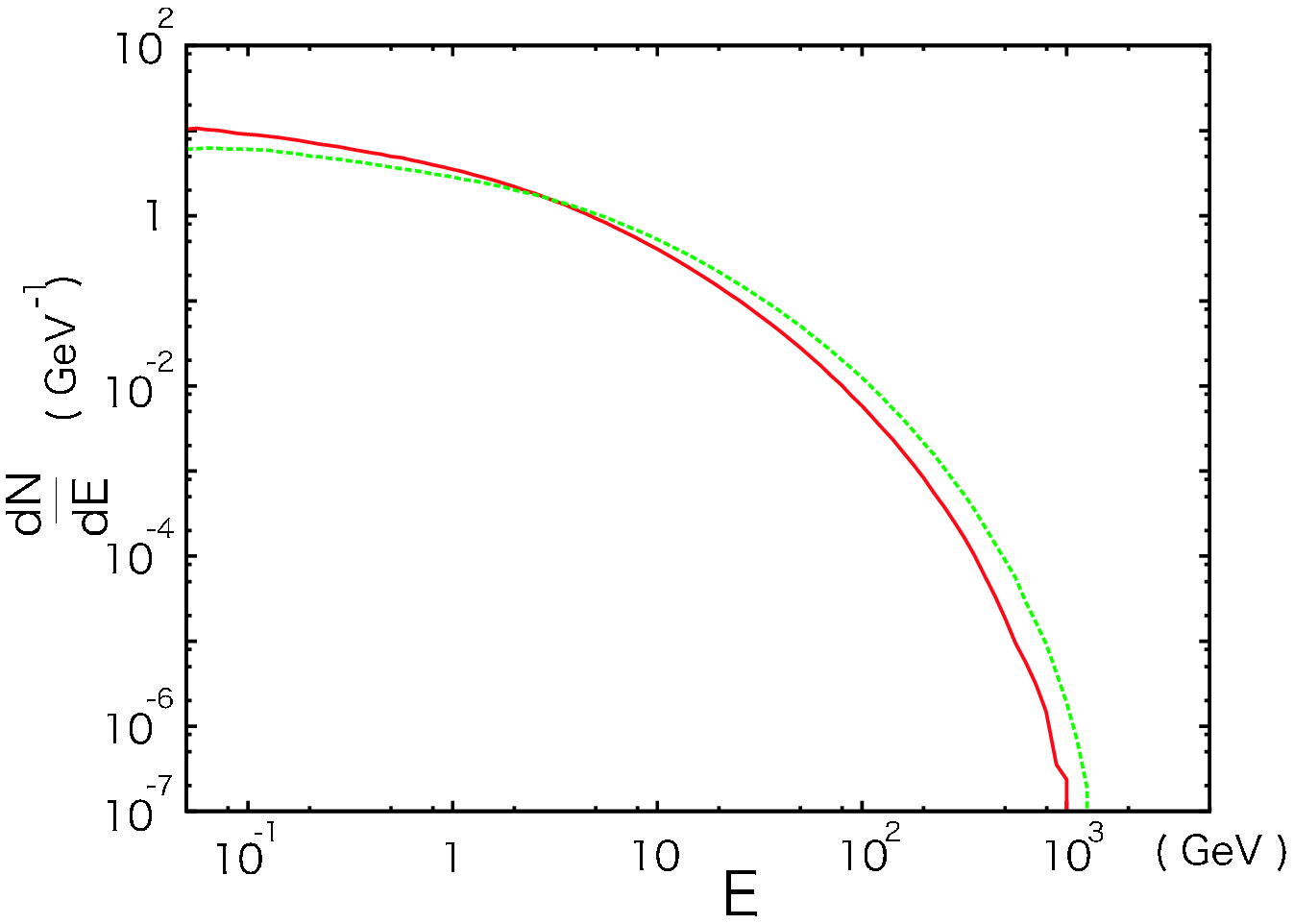}
                \caption{Same as Fig.\ref{fig:decay_m=1e2} except 
                $m_X=10^4{\rm GeV}$.}
        \label{fig:decay_m=1e4}
        \end{center}
\end{figure}

\begin{figure}[htbp]
        \begin{center}
                \includegraphics[width=0.8\linewidth]{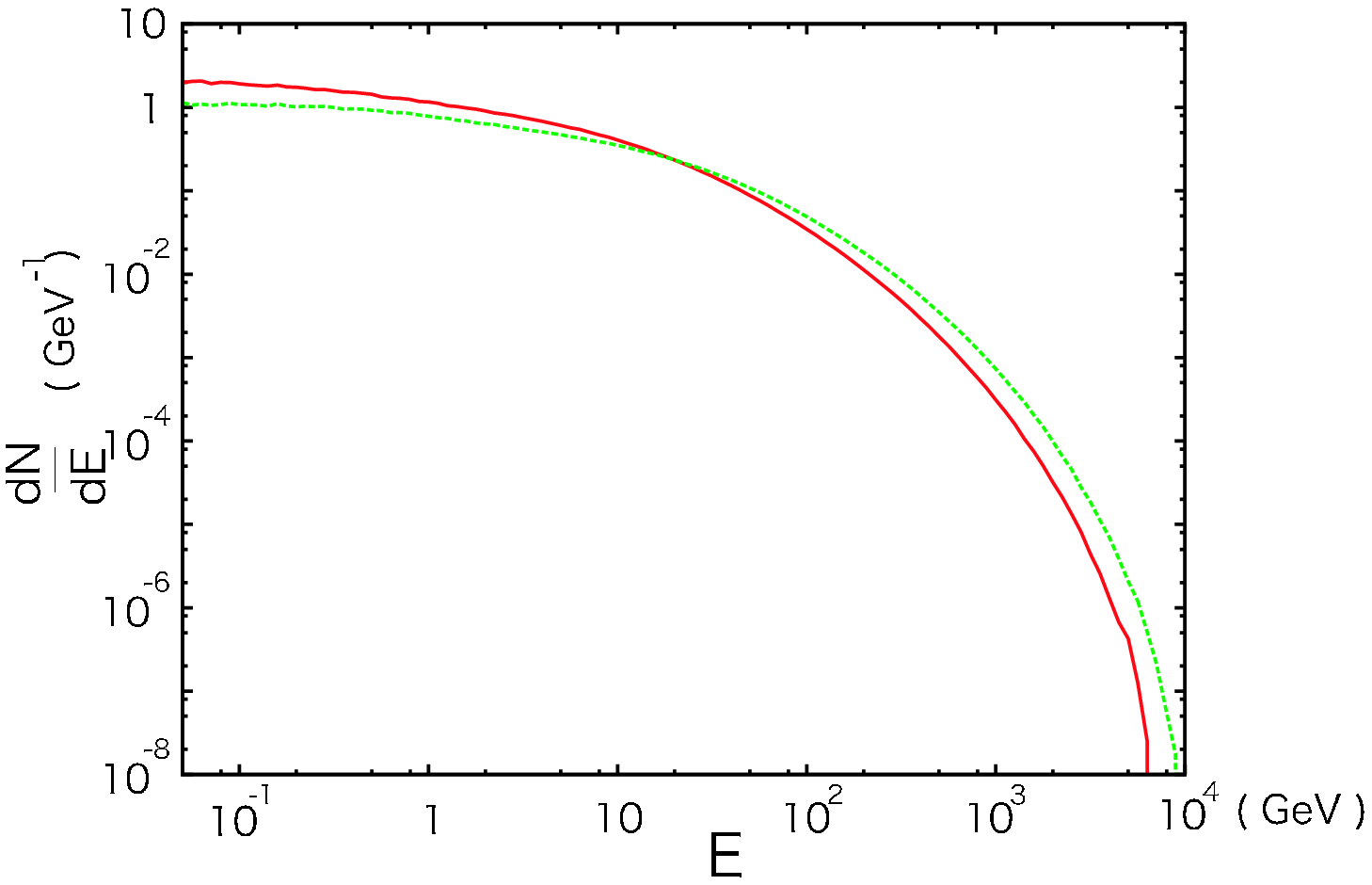}
                \caption{Same as Fig.\ref{fig:decay_m=1e2} except 
                $m_X=10^5{\rm GeV}$.}
        \label{fig:decay_m=1e5}
        \end{center}
\end{figure}

In order to study the effects of three- and four-body decay processes,
it is important to obtain the spectra of quarks and leptons emitted by
the decay of $X$, which depends on the model.  In our analysis, we
use, up to normalization, those obtained in the case where $X$ is the
sneutrino and $Y$ is the gravitino; in such a case, the sneutrino may
decay into the gravitino, neutrino (or charged lepton), and $Z^{(*)}$
(or $W^{(*)}$), and the produced (real or virtual) weak bosons
subsequently decay into quark- or lepton-pair.  (Here, $Z^{*}$ and
$W^{*}$ denote virtual weak bosons.)  In order to perform our analysis
as model-independent as possible, we treat that the branching ratio
for three- and four-body processes is a free parameter: we define
branching ratio $B_X$ as
\begin{eqnarray}
  B_X\equiv\frac{\Gamma(X\to{\rm 3~body})+\Gamma(X\to{\rm 4~body})}
  {\Gamma(X\to{\rm all})},
\end{eqnarray}
where $\Gamma$ is decay width.  For the study of the effects of
photo-dissociation processes and CMB spectral distortion, we also
calculate averaged ``visible energy'' emitted from one $X$:
\begin{eqnarray}
        E_{\rm vis}=B_X\langle E_{\rm vis}\rangle,
\end{eqnarray}
where $\langle E_{\rm vis}\rangle$ are averaged energy carried away by
charged particles and photons in three and four-decay modes.  The
energy distributions of photons, neutrinos, leptons and nucleons
produced by the decay of $X$ are calculated by means of Monte Carlo
simulations with the PYTHIA package \cite{Sjostrand:2000wi}.  In
Figs.\ \ref{fig:decay_m=1e2}-\ref{fig:decay_m=1e5}, we plot the
spectra of photons and leptons ($e^+ + e^-$), respectively.  From
these distributions we calculate the averaged energy and obtain
$\langle E_{\rm vis}\rangle = 25.3, 146, 821$ and $5630\ {\rm GeV}$
for $m_X=10^2,10^3,10^4$ and $10^5\ {\rm GeV}$, respectively.  Here,
$m_X$ is the mass of $X$.  $E_{\rm vis}$ is much smaller than $m_X$
since we are interested in the case where $B_X\ll 1$.


Photo-dissociation processes are induced by energetic photons in
electromagnetic shower which is caused by charged particles and/or
photons emitted from $X$.  With given background temperature, the
distribution function of energetic photons depends on total amount of
energy injected by particles with electromagnetic interaction, and is
insensitive to the shape of energy spectrum of primary particles.  Thus,
once $E_{\rm vis}$ is obtained, the energy distribution of energetic
photons in the electromagnetic shower can be obtained.  Then,
photo-dissociation rates are obtained by convoluting energy distribution
function and cross sections of photo-dissociation reactions.  For
details of our treatment of photo-dissociation processes,
see~\cite{KMrad}.

For the study of hadro-dissociation processes, it is necessary to
obtain energy distributions of (primary) hadrons which are produced
after the hadronization of quarks emitted from $X$.  We have
calculated the spectra of $p$ and $n$ by using PYTHIA.  These hadrons
cause hadronic shower and induced hadro-dissociation processes.  In
our analysis, in addition, we have also calculated the number of
charged pions produced by the decay of $X$.  Such charged pions,
protons and neutrons become the source of $p\leftrightarrow n$
conversion process, which changes the number of ${}^4{\rm
  He}$~\cite{Reno:1987qw,Kohri:2001jx}.\footnote
{Here we have neglected the effects of Kaons
to the $p\leftrightarrow n$ conversion, according to the
discussion in~\cite{KKM04}.}
Once the spectra of hadrons are
obtained, effects of hadro-dissociation  and $p\leftrightarrow n$
conversion are studied with the procedure given in \cite{KKM04}.

\subsection{Numerical Results}

In our analysis, we have followed the evolutions of the number densities
of the light elements.  For this purpose, we have modified the Kawano
code \cite{Kawano:1992ua} including photo- and hadro-dissociation
processes.
As observational constraints on the primordial
abundances of light elements, we adopt those used in
\cite{Kanzaki:2006hm} except for $Y_{\rm p}$ and $(n_{\rm ^3He}/n_{\rm
D})_{\rm p}$:
\begin{eqnarray}
  (n_{\rm D}/n_{\rm H})_{\rm p} &=& (2.82\pm 0.26) \times 10^{-5},
  \\
   (n_{\rm ^3He}/n_{\rm D})_{\rm p} &<& 0.83+0.27,
  \\
   Y_{\rm p} &=& 0.2516 \pm 0.0040,
  \\
  \log_{10}(n_{\rm ^7Li}/n_{\rm H})_{\rm p} &=& -9.63 \pm 0.06 \pm 0.3.
  \\
  (n_{\rm ^6Li}/n_{\rm ^7Li})_{\rm p} &<& 0.046\pm 0.022 + 0.084,
  \label{Li6/Li7}
\end{eqnarray}
where the subscript ``p'' is for primordial value (just after BBN),
and $Y_{\rm p}$ is the primordial mass fraction of ${\rm ^4He}$. For
the center value of $Y_{\rm p}$, we have adopted the value
reported in \cite{Izotov:2007ed} in which the authors used new data of
HeI emissivities,\footnote
{See also the other recent value of $Y_{\rm p}$ reported in
\cite{Peimbert:2007vm} where the authors adopted larger errors
(0.0028) than that of \cite{Izotov:2007ed}.}
and conservatively added a larger error (= $0.0040$) as discussed in
\cite{Fukugita:2006xy}.  For $(n_{\rm ^3He}/n_{\rm D})_{\rm p}$, we
have adopted most newly-reported values of D and $^{3}$He abundances
observed in protosolar clouds~\cite{GG03}, $(n_{\rm ^3He}/n_{\rm
  H})_{\rm PSC} = (1.66 \pm 0.06) \times 10^{-5}$ and $(n_{\rm
  D}/n_{\rm H})_{\rm PSC} = (2.00 \pm 0.35) \times 10^{-5}$, where the
subscript ``PSC'' means a value in the protosolar cloud.  (For the
importance of the upper bounds on $(n_{\rm ^3He}/n_{\rm D})_{\rm p}$,
see~\cite{Sigl:1995kk,KKM04}.)

We parameterize the primordial abundance of $X$ by yield variable $Y_X$
which is defined as the ratio of number density and 
total entropy density at ($t \ll \tau_X$);
\begin{eqnarray}
        Y_X\equiv\left[\frac{n_X}{s}\right]_{t\ll \tau_X},
\end{eqnarray}  
where $\tau_X$ is the lifetime of $X$.  If $Y_X$ is too large,
abundances of light elements are too much affected to be consistent
with the observations.  Thus, we can derive upper bound on $Y_X$.


\begin{figure}[t]
        \begin{center}
                \includegraphics[width=0.7\linewidth]{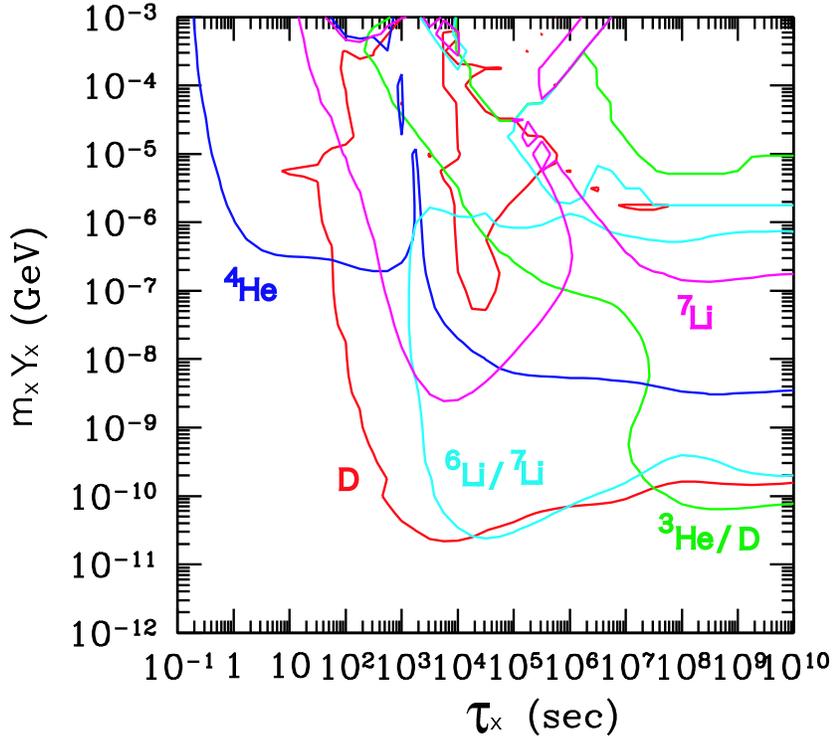}
        \caption{BBN constraints on $\tau_X$ vs. $m_XY_X$ plane.
        Here, we take $m_X=100{\rm GeV}$ and $B_X=10^{-3}$.}
        \label{fig:BbnNu_E2Bm3}
        \end{center}
\end{figure}

\begin{figure}[thbp]
        \begin{center}
                \includegraphics[width=0.7\linewidth]{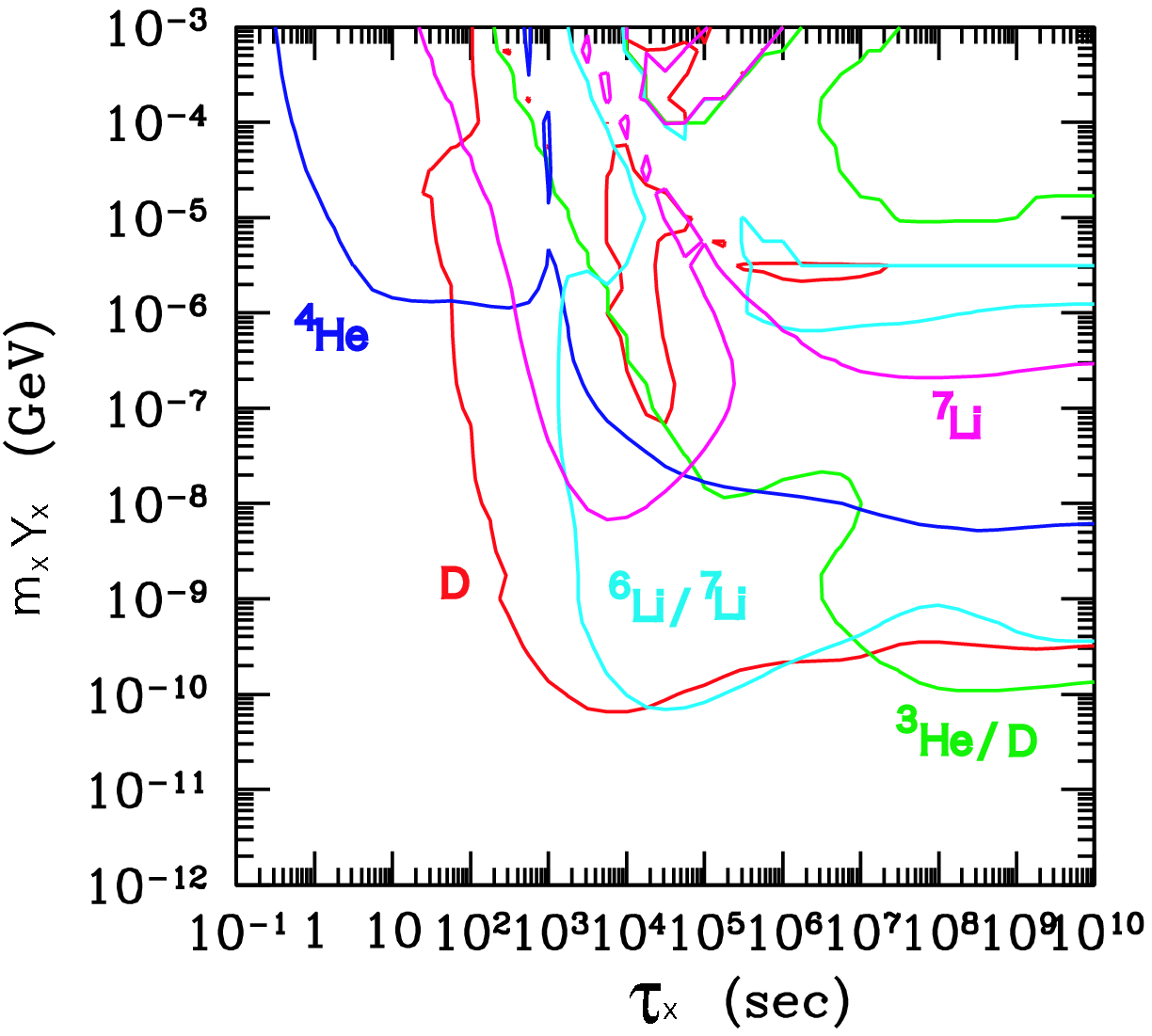}
        \caption{BBN constraints on $\tau_X$ vs. $m_XY_X$ plane.
        Here, we take $m_X=1{\rm TeV}$ and $B_X=10^{-3}$.}
        \label{fig:BbnNu_E3Bm3}
        \end{center}
\end{figure}

\begin{figure}[thbp]
        \begin{center}
                \includegraphics[width=0.7\linewidth]{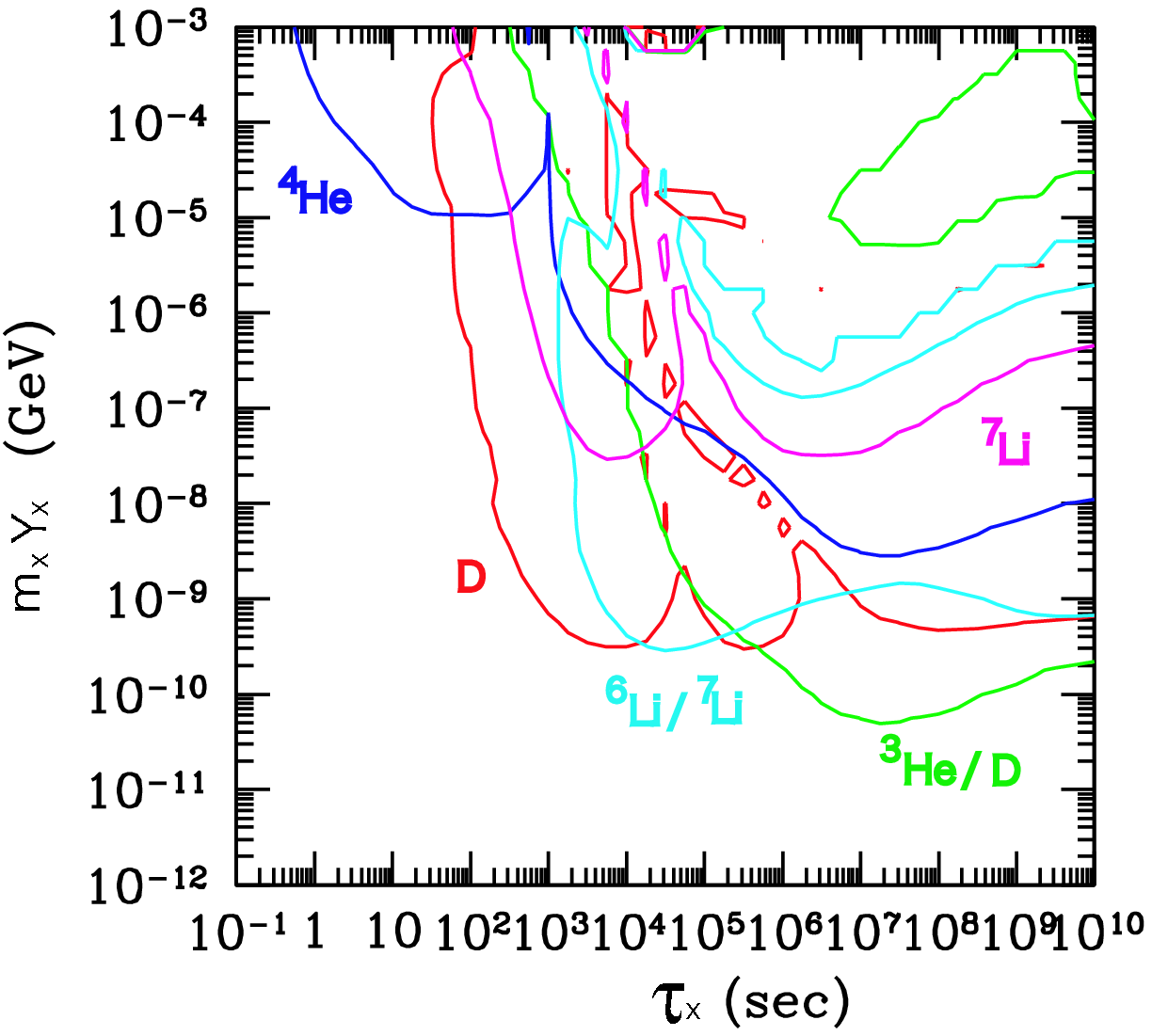}
        \caption{BBN constraints on $\tau_X$ vs. $m_XY_X$ plane.
        Here, we take $m_X=10{\rm TeV}$ and $B_X=10^{-3}$.}
        \label{fig:BbnNu_E4Bm3}
        \end{center}
\end{figure}

\begin{figure}[thbp]
        \begin{center}
                \includegraphics[width=0.7\linewidth]{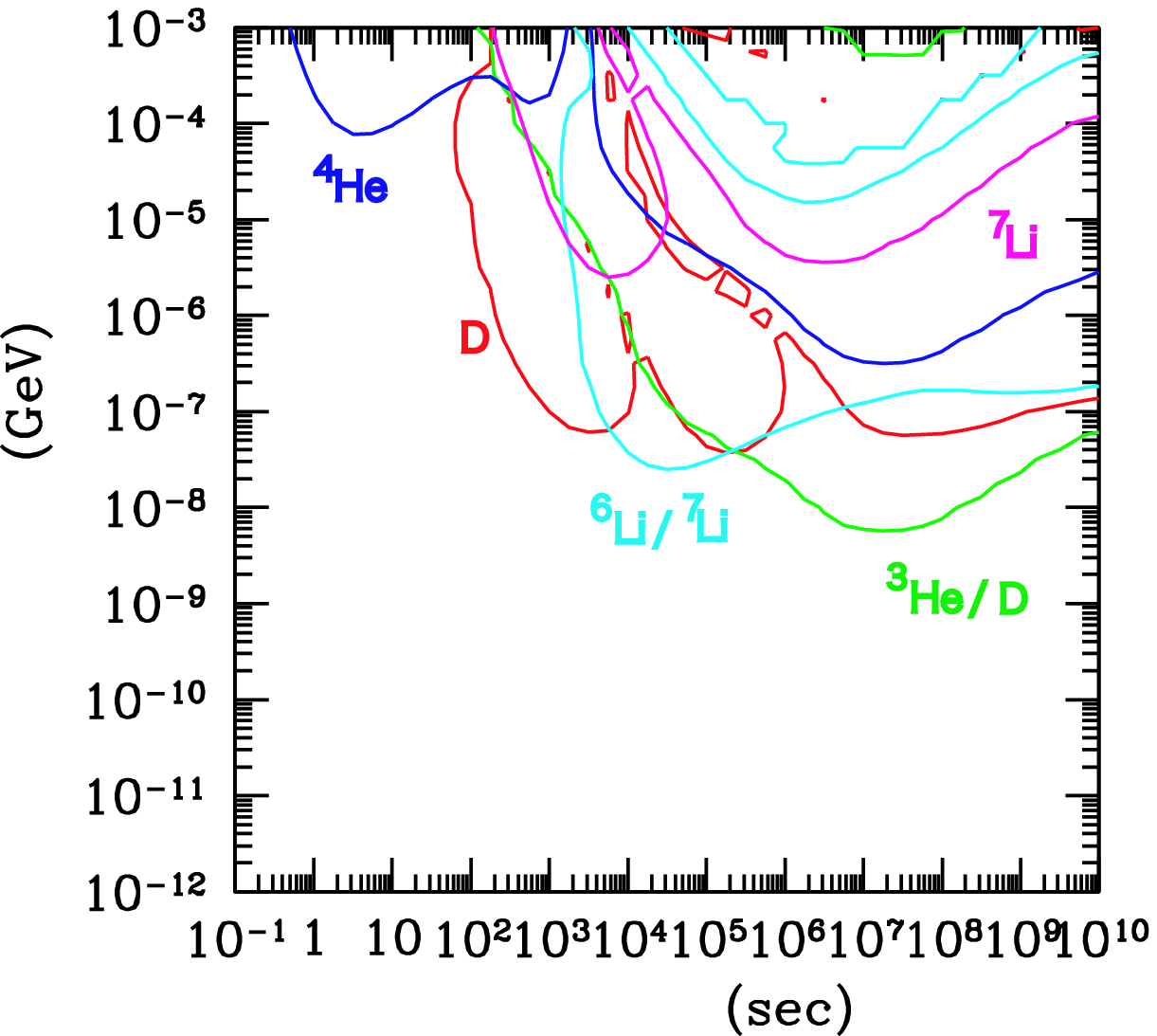}
        \caption{BBN constraints on $\tau_X$ vs. $m_XY_X$ plane.
        Here, we take $m_X=100{\rm GeV}$ and $B_X=10^{-6}$.}
        \label{fig:BbnNu_E2Bm6}
        \end{center}
\end{figure}

\begin{figure}[thbp]
        \begin{center}
                \includegraphics[width=0.7\linewidth]{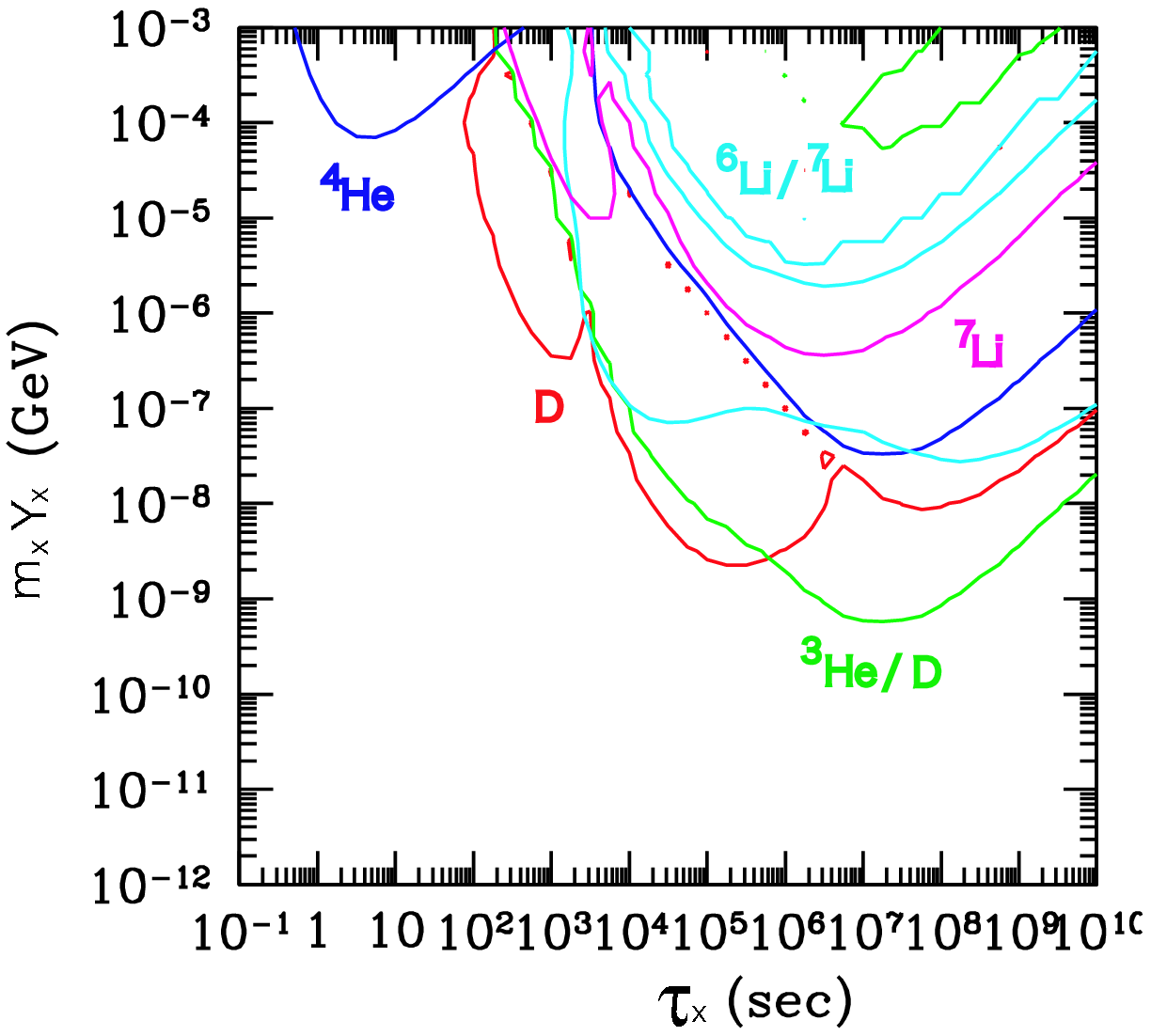}
        \caption{BBN constraints on $\tau_X$ vs. $m_XY_X$ plane.
        Here, we take $m_X=1{\rm TeV}$ and $B_X=10^{-6}$.}
        \label{fig:BbnNu_E3Bm6}
        \end{center}
\end{figure}

\begin{figure}[thbp]
        \begin{center}
                \includegraphics[width=0.7\linewidth]{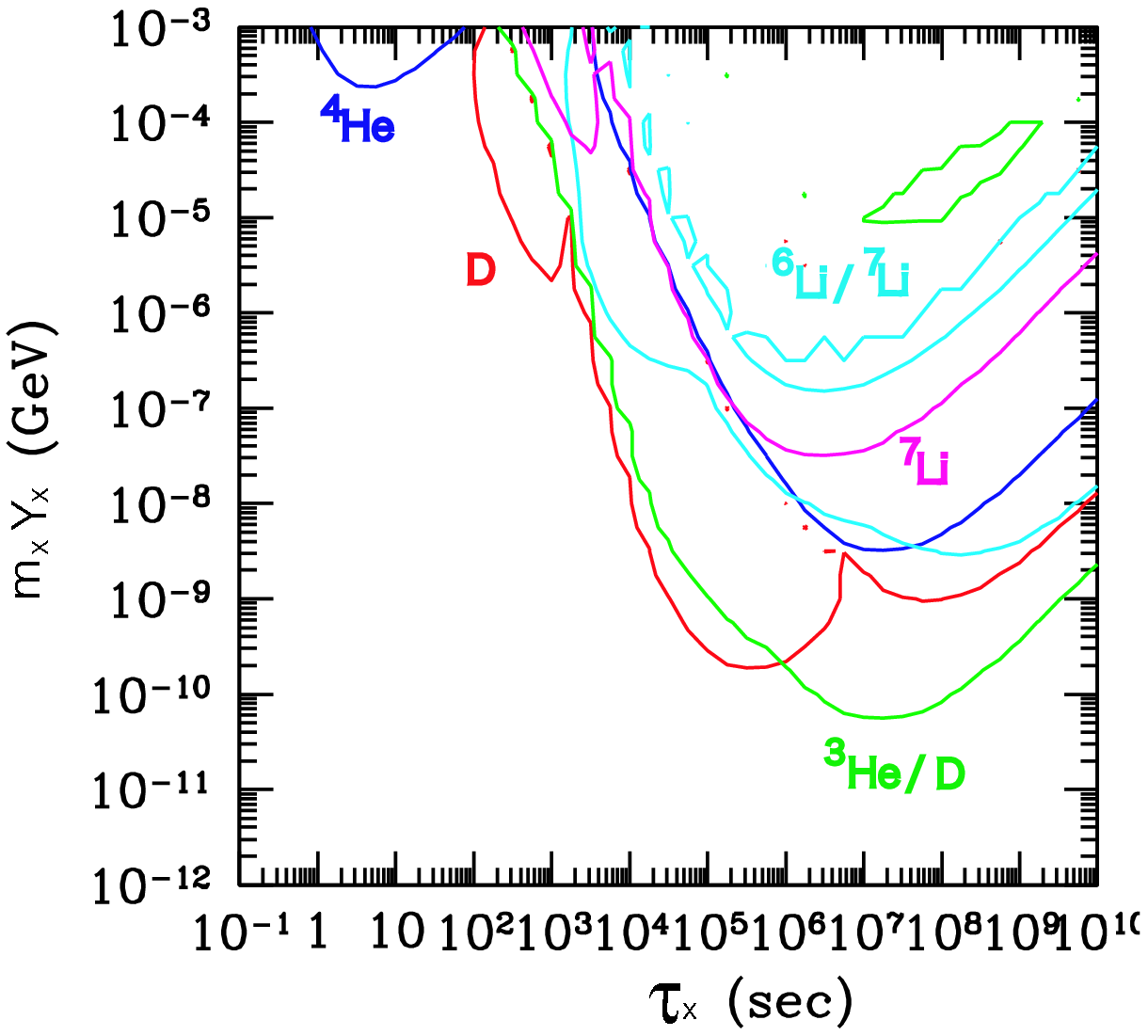}
        \caption{BBN constraints on $\tau_X$ vs. $m_XY_X$ plane.
        Here, we take $m_X=10{\rm TeV}$ and $B_X=10^{-6}$.}
        \label{fig:BbnNu_E4Bm6}
        \end{center}
\end{figure}


In Figs.\ \ref{fig:BbnNu_E2Bm3}-\ref{fig:BbnNu_E4Bm6}, we show BBN
constraints on $\tau_X$ and $m_XY_X$ for $B_X=10^{-3}$ and
$B_X=10^{-6}$; we found that constraints with $B_X=0$ is almost same as
that with $B_X=10^{-6}$.

As one can see from these figures, the most important constraint comes
from overproduction of ${}^4{\rm He}$ when $\tau_X\lesssim10^2{\rm
  sec}$.  Since protons are more abundant than neutrons, a significant
amount of proton may be converted to neutron through nucleus-pion
interaction (and nucleon-nuclen interactions for relatively large
$B_{X}$) and consequently ${}^4{\rm He}$ is overproduced.  When
$10^2{\rm sec}\lesssim \tau_X \lesssim 10^7{\rm sec}$, the background
${}^4{\rm He}$ (which we call $\alpha_{\rm BG}$) is effectively
dissociated by the energetic hadrons produced in the hadronic shower.
In this case, overproduction of D may occur as a result of
hadro-dissociation of $\alpha_{\rm BG}$.  In addition, energetic T and
${}^3{\rm He}$ are also produced and they synthesize $^6$Li through
the ${}^6{\rm Li}$ via ${\rm T}+\alpha_{\rm BG}\to {}^6{\rm Li}+n$ and
${}^3{\rm He}+\alpha_{\rm BG}\to {}^6{\rm Li}+p$.  When
$\tau_X\gtrsim10^7{\rm sec}$, the energetic hadrons are stopped by the
scattering processes with background electrons, and hence the effects
of hadro-dissociation become less efficient than those of the
photo-dissociation.  In particular, the energetic photons produced in
the electromagnetic shower destroy $\alpha_{\rm BG}$.  In this case,
overproduction of D and ${}^3{\rm He}$ occurs as a result of
photo-dissociation of $\alpha_{\rm BG}$.

The constraints on $m_X Y_X$ depend on $m_X$ in a non-trivial way.  The
contraint coming from the photo-dissociation caused by the two-body
decay becomes stringent as $m_X$ increases because neutrinos emitted in
two-body decay have higher energy and scatter off the background
neutrinos with larger rate. On the other hand, the rates of the
hadro-dissociation and photo-disocciation caused by the three- and
four-body decay depend only on $E_{\rm vis} Y_X$ and hence the
constraints on $m_XY_X$ become slightly milder as $m_X$
increses.  The constraint from $n \leftrightarrow p$ conversion becomes
weaker as $m_X$ increases. This is because the charged pion production
is roughly determined by $Y_X$ for the three- and four-decay.

So far, we have shown the results for the case where the primary
neutrino is electron-type.  However, we have checked that the BBN
constraints are almost unchanged even if $X$ decays into muon- or
tau-neutrino (and $Y$). 

\section{CMB Constraints}
\label{sec:cmb}

CMB also imposes constraints on the decays of $X$.
COBE observations show that CMB spectrum is almost perfect 
blackbody~\cite{FIRAS}.
Therefore any exotic energy injections that cause distortions 
in the spectrum of CMB are stringently 
constrained~\cite{Silk:1983hj,Kawasaki:1985ff,Hu:1993gc}.
When the photons are emitted before redshift  $z\sim 10^7$, 
they are thermalized by Compton scattering, double Compton 
scattering and bremsstrahlung, and no spectral distortion 
takes place.
However, at  $z \lesssim 10^7$ only Compton scattering is efficient.
Since the Compton scattering does not change the total number of 
photons, the resultant spectrum becomes a Bose-Einstein 
distribution with a finite chemical potential $\mu$, 
regardless of the detail of the injection.
For the case of massive particle decay, 
the number density of injected photons is negligible compared 
with that in the background.
Therefore, the spectral distortions are determined by the fraction 
of the energy release, $\Delta\rho_\gamma/\rho_\gamma$.  
Then, for small $\Delta\rho_\gamma/\rho_\gamma$ 
the chemical potential of the photon spectrum is given by
\begin{eqnarray}
 \mu 
 \simeq\frac{1}{0.714}\frac{\Delta\rho_\gamma}{\rho_\gamma}.
        \label{mu_energy_relation}
\end{eqnarray}  

As we have discussed, there are two types of processes which contribute
to the electromagnetic energy injection ($\Delta\rho_\gamma$) in the
present scenario.  One is the two-body decay process: the high energy
primary neutrinos emitted by $X$ scatter off the background neutrinos
and create charged leptons whose energy is finally converted to the
energy of radiation.  The other is the three- and four-body decay
processes by which quarks and charged leptons are produced.  In our
analysis, we have taken into account both of these contributions.


\subsection{Three and four-body Decay}

There are two quantities which are required to calculate 
the electromagnetic energy injection:
branching ratio $B_X$ and the averaged energy $\langle E_{\rm vis}\rangle$.

We follow the treatment of~\cite{Hu:1993gc} for the case of 
three and four-body decays.
Using the fact that $X$
decays exponentially in time with lifetime 
$\tau_X$, we obtain
\begin{eqnarray}
  \frac{\Delta\rho_\gamma}{\rho_\gamma}
  =\frac{\langle E_{\rm vis}\rangle}{2.701T(t_{\rm eff})}
  \frac{n_X}{n_\gamma}B_X ,
  \label{energy_ratio}
\end{eqnarray}
where $T(t)$ is the CMB temperature  and 
$n_X$ is the number density of $X$ before decay.
Here $t_{\rm eff} = \left[\Gamma(1+\beta)\right]^{1/\beta}\tau_X$
for time-temperature relation $T\propto t^{-\beta}$,
where $\Gamma$ is the gamma function. 

From Eqs.~(\ref{mu_energy_relation}) and (\ref{energy_ratio}), 
we find that the chemical potential is given by
\begin{eqnarray}
        \mu\simeq4.00\times10^2
        \left(\frac{\tau_X}{1{\rm sec}}\right)^{1/2}
        \left(\frac{\langle E_{\rm vis}\rangle}{1{\rm GeV}}\right)
        B_X\frac{n_X}{n_\gamma}.
        \label{mu_decay}
\end{eqnarray}
We have assumed here that we are in the radiation dominated 
epoch where $T\propto t^{-1/2}$.
Note that, however, photon number changing processes 
(double Compton scattering and bremsstrahlung) become 
increasingly efficient as the photon frequency decreases.
This means the spectrum becomes blackbody at low frequencies.
The photons with low frequencies produced by the photon-number
changing processes are transferred to higher frequencies 
by inverse Compton scattering and the chemical potential 
decreases in time.  
For a low $\Omega_b h^2$ universe suggested by the
BBN~\cite{Steigman:2005uz} and WMAP~\cite{Spergel:2006hy}
(where, in this paper, $\Omega_b$ denotes the density parameter
of baryon, and $h$ is the Hubble constant in units of
$100\ {\rm km/sec/Mpc}$), double
Compton scattering dominates the thermalization process.  The chemical
potential produced at $t=t_h$ is blurred out at an exponential rate and
the present value is given by~\cite{Danese;1982}
\begin{eqnarray}
        \mu_0\equiv\mu(t_0)=\mu(t_h)\exp(-(t_{\rm DC}/t_h)^{5/4})
        \label{mu_evolution}
\end{eqnarray}
with
\begin{eqnarray}
        t_{\rm DC}=6.81\times10^6
        \left(\frac{\Omega_b h^2}{0.0223}\right)^{4/5}
        \left(1-\frac{Y_p}{2}\right)^{4/5} {\rm sec}.
\end{eqnarray}
Combining Eq.~(\ref{mu_decay})
with Eq.~(\ref{mu_evolution}), we find that the chemical potential today
is given by~\cite{Hu:1993gc}
\begin{eqnarray}
        \mu_0\simeq4.00\times10^2
        \left(\frac{\tau_X}{1{\rm sec}}\right)^{1/2}
        \exp(-(t_{\rm DC}/\tau_X)^{5/4})
        \left(\frac{\langle E_{\rm vis}\rangle}{1{\rm GeV}}\right)
        B_X\frac{n_X}{n_\gamma}.
\end{eqnarray}

For late energy injection $(z \lesssim 10^5 )$, Compton scattering can
no longer establish the Bose-Einstein distribution.  In this case, the
spectrum can be described by the Compton $y$-parameter which is defined
by
\begin{equation}
    y = \int dt \frac{T_e -T}{m_e}n_e\sigma_T,
\end{equation}
where $m_e$ and $T_e$ are the number density and temperature of 
electrons and $\sigma_T$ is Thomson cross section. 
Then the energy injection is related to $y$ as
$\Delta\rho_\gamma/\rho_\gamma=4y$.
Here, we define $z_K$ as the redshift at which 
the time scale for energy exchange through Compton scatterings 
is equal to the Hubble time;
\begin{eqnarray}
        z_K\simeq4.77\times10^4
        \left(\frac{\Omega_b h^2}{0.0223}\right)^{-1/2}
        \left(1-\frac{Y_p}{2}\right)^{-1/2}.
\end{eqnarray}
The spectrum can be described by chemical potential $\mu$ 
for energy injection at $z>z_K$ 
and by Compton $y$-parameter  for energy injection at $z<z_K$.
Here we take $\Omega_b h^2=0.0223$~\cite{Spergel:2006hy}, 
$|\mu|<9\times10^{-5}$ and 
$|y|<1.2\times10^{-5}$ as observational
limits by COBE~\cite{Fixsen:1996nj,Hagiwara:2002fs}.

\subsection{Two-body Decay}

\begin{figure}[t]
        \begin{center}
                \includegraphics[width=0.8\linewidth]{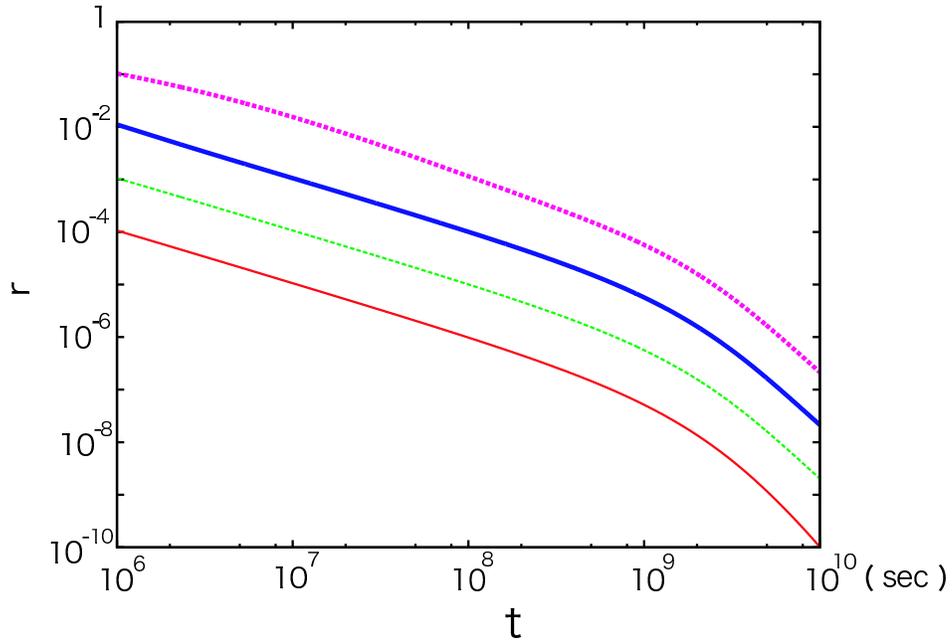}
                \caption{The time evolution of $r$ with $m_X=10^2{\rm GeV}$ 
                (thin solid line), $10^3{\rm GeV}$ (thin dashed line), 
                $10^4{\rm GeV}$ (thick solid line) and 
        $10^5{\rm GeV}$ (thick dashed line).
        We take $\tau_X = 10^9{\rm sec}$.}
        \label{fig:ratio1}
        \end{center}
\end{figure}

\begin{figure}[t]
  \begin{center}
    \includegraphics[width=0.8\linewidth]{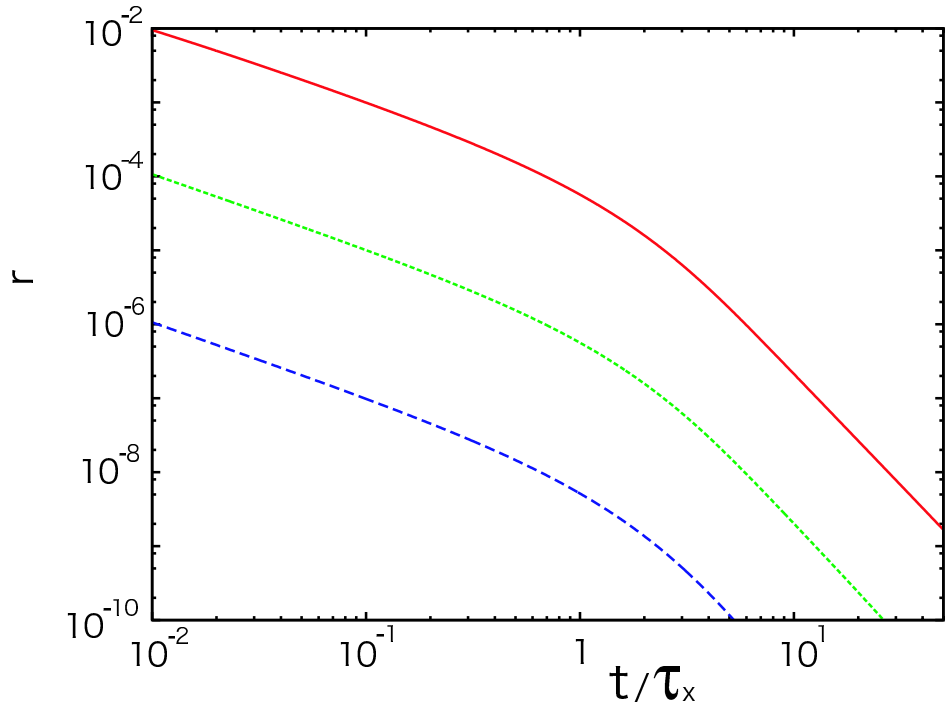}
    \caption{The time evolution of $r$ with $\tau_X=10^7{\rm sec}$ 
      (solid line), $10^9{\rm sec}$ (dotted line), 
      $10^{11}{\rm sec}$ (dashed line). 
      We take $m_X = 10^3{\rm GeV}$.}
    \label{fig:ratio2}
  \end{center}
\end{figure}

In addition, the effect of charged-lepton productions 
through  
scattering of high energy neutrinos off background neutrinos 
(see Eqs.\ (\ref{nunu2ee}) -- (\ref{nunu2emu}))
should be taken into account.
The amount of energy which is converted to the background photons 
is estimated by
\begin{eqnarray}
        \frac{\Delta\rho_\gamma}{\rho_\gamma} 
        = \int^\infty_0 dt \frac{1}{\rho_\gamma}\frac{dE_l}{dt}
        = \frac{m_X}{2}\frac{n_X}{n_\gamma}
        \int^\infty_0 dt~ \frac{1}{2.701T}
        \frac{1}{\tau_X}r(t,m_x,\tau_X),
        \label{neutrino_inj}
\end{eqnarray}  
where $dE_l/dt$ is the energy density which is converted to 
charged lepton and finally photon 
through neutrino scattering per unit time 
and 
\begin{eqnarray}
 r \equiv \left(\frac{m_X n_X}{2\tau_X}\right)^{-1} \frac{dE_l}{dt}.
\end{eqnarray}
Notice that $r$ represents the ratio of the radiative energy 
injection per $X$ decay to the $X$ mass.
The details of calculation are found  
in~\cite{Kawasaki:1994bs,Moroi:1995fs}.
In Figs.\ \ref{fig:ratio1} and \ref{fig:ratio2}, we show the time 
evolution of $r$.
Fig.~\ref{fig:ratio1} shows that the ratio $r$ increases 
with $m_X$.
This is because higher energy neutrino has larger cross section
for scattering off the background neutrino 
and is also easy to exceed the 
threshold energy for lepton pair creations.
The similar logic applies to Fig.~\ref{fig:ratio2}.
In this case, background neutrino energy increases with the 
decrease 
of $\tau_X$, so that the ratio $r$ increases. 


\subsection{Constraint from CMB}

\begin{figure}[t]
  \begin{center}
    \includegraphics[width=0.8\linewidth]{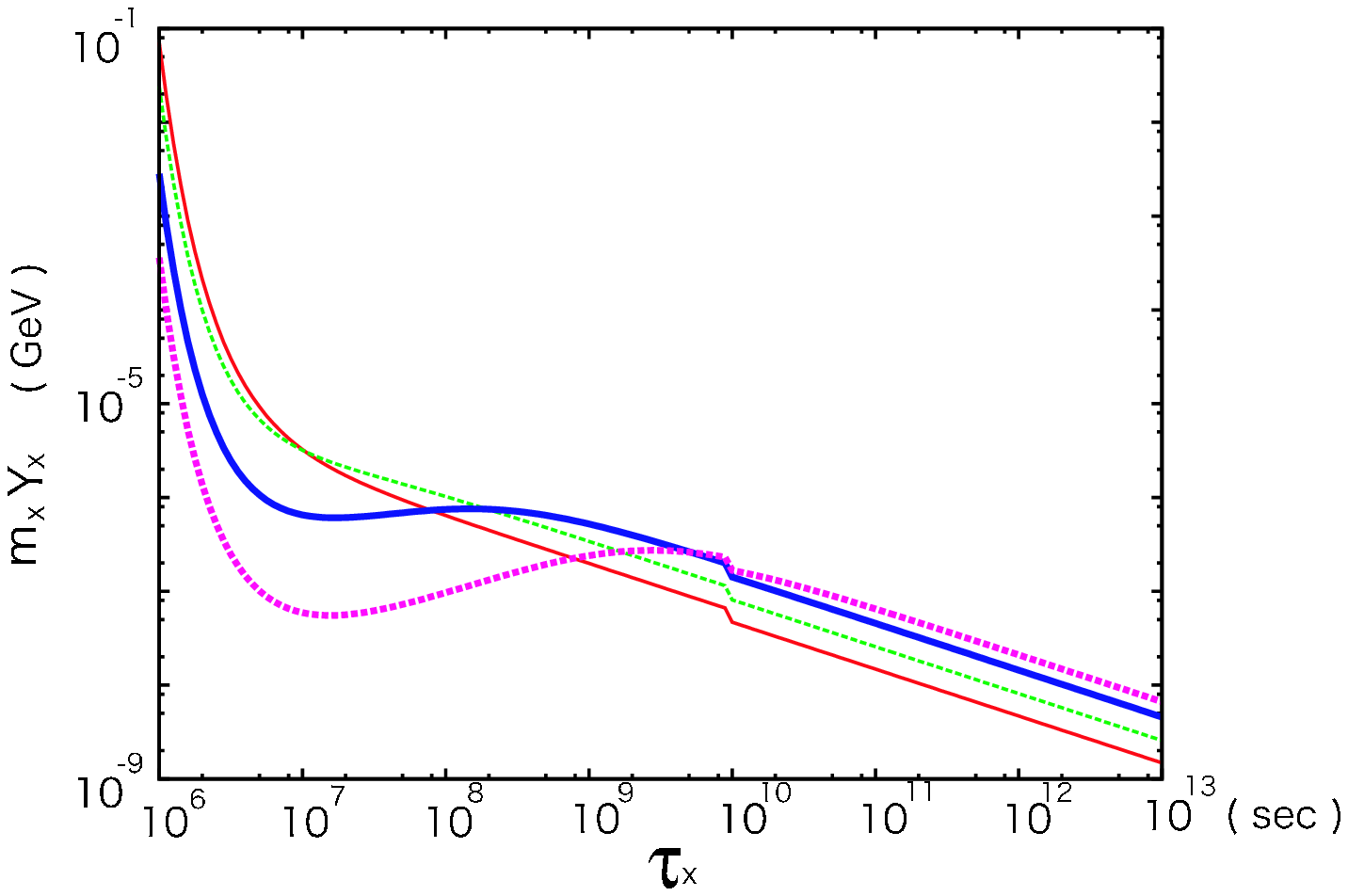}
    \caption{CMB constraints on $m_XY_X$ with $B_X=10^{-3}$.
      From upper to lower the lines represent 
      the upper bound of $m_XY_X$ when $m_X=10^2{\rm GeV}$ 
      (thin solid line), $10^3{\rm GeV}$ (thin dotted line), 
      $10^4{\rm GeV}$ (thick solid line) 
      and $10^5{\rm GeV}$ (thick dotted line).}
    \label{fig:cmb_b3}
  \end{center}
\end{figure}

\begin{figure}[thbp]
  \begin{center}
                \includegraphics[width=0.8\linewidth]{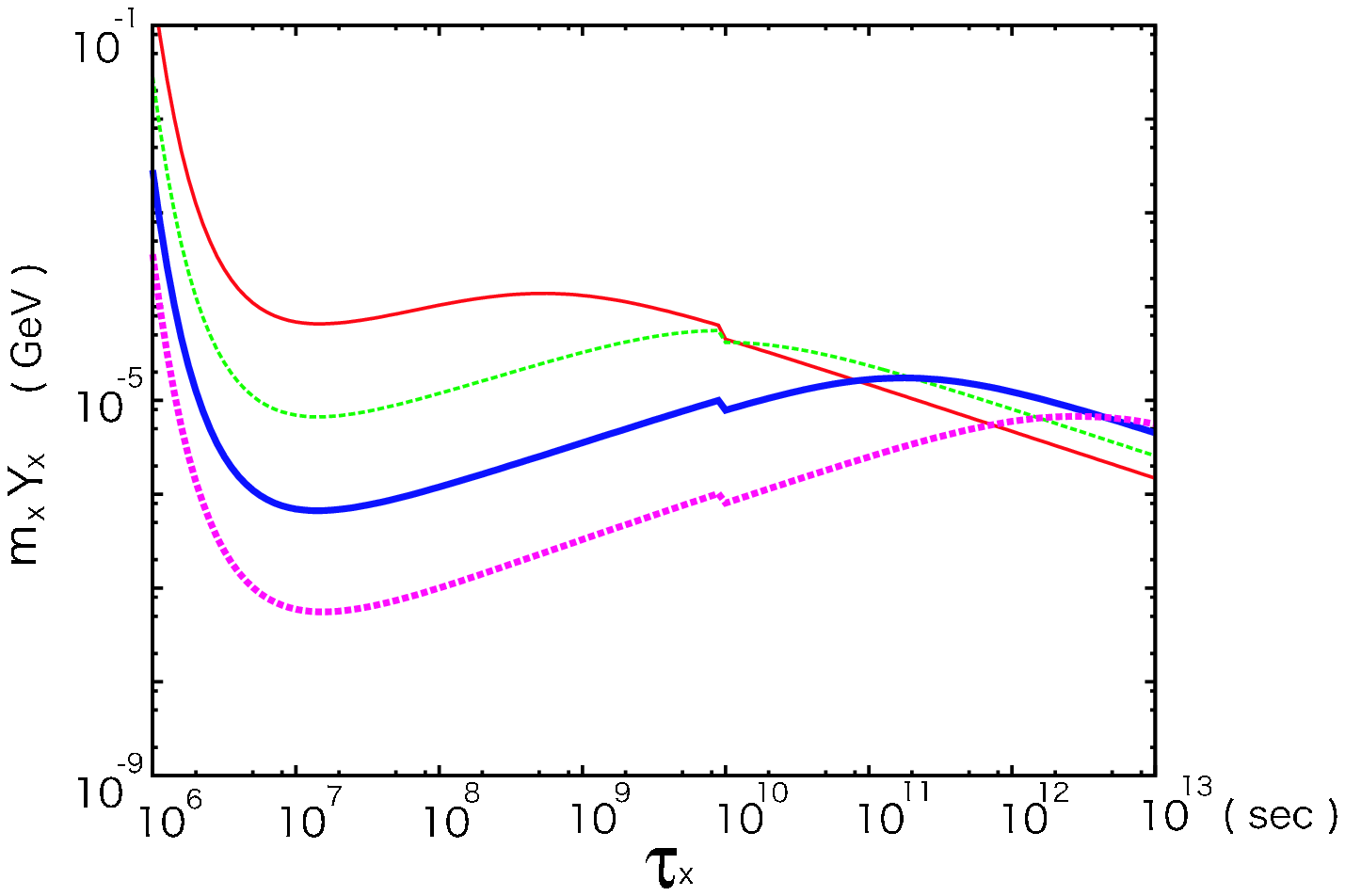}
                \caption{Same as Fig.~\ref{fig:cmb_b3} except $B_X=10^{-6}$.} 
        \label{fig:cmb_b6}
        \end{center}
\end{figure}

As mentioned earlier, the spectral distortions are determined 
by $\Delta\rho_\gamma/\rho_\gamma$ which is the sum of 
Eqs.~(\ref{energy_ratio}) and (\ref{neutrino_inj}).
In Figs.~\ref{fig:cmb_b3} and \ref{fig:cmb_b6}, 
we show the upper bounds of $m_XY_X$ 
taking account of neutrino-neutrino scattering.


When the lifetime is short and the energy of background neutrino 
is sufficiently high, $\mu$ is determined by neutrino-neutrino 
scattering, hence $r$.
Therefore, the constraints become severer as $m_X$ becomes larger.
When lifetime is long, $\mu$ is determined by 
three and four-body decay of $X$, hence 
$\langle E_{\rm vis}\rangle/m_X$.
As already mentioned, $\langle E_{\rm vis}\rangle/m_X$ 
becomes smaller with larger $m_X$.
Then, the constraints become severer as $m_X$
becomes smaller.  

\begin{figure}[thbp]
        \begin{center}
                \includegraphics[width=0.8\linewidth]{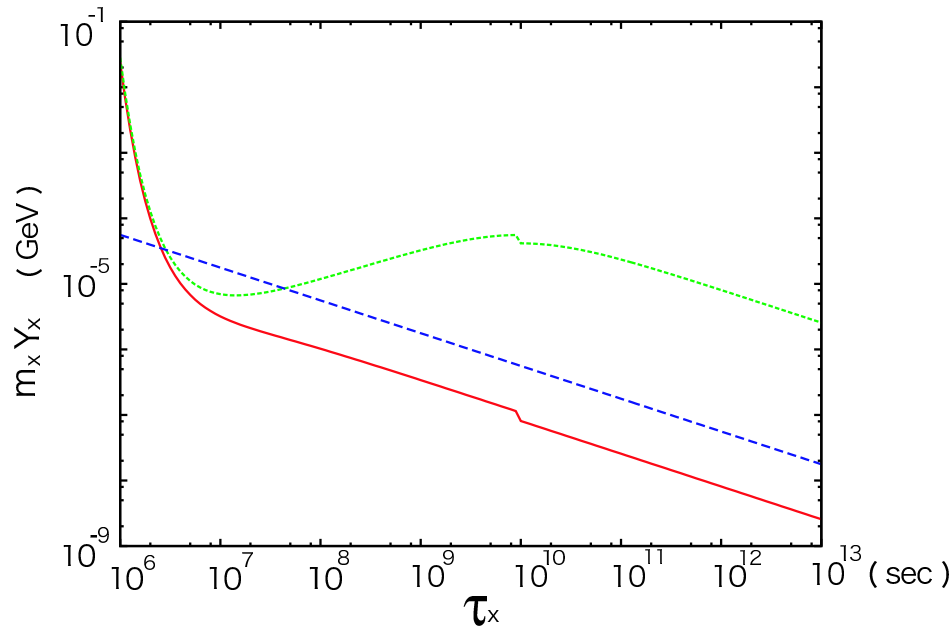}
                \caption{Solid and dotted lines represent the CMB 
                constraints on $m_X Y_X$ 
                with $m_X=1~\mathrm{TeV}$ when $B_X = 10^{-3}$ and 
                $10^{-6}$ respectively.
                Dashed line represents the constraints from increases of
                the total relativistic energy.}
        \label{fig:cmb} \end{center}
\end{figure}

So far, we have focused only on photon energy injection.  However, the
emitted ultra-relativistic particles (neutrino and $Y$) contribute to
the total relativistic energy and could lead to a more stringent
constraint than that from spectral distortion, when branching ratio is
sufficiently small.  Before recombination, the CMB angular power
spectrum is sensitive to the change of the total relativistic energy
through the early integrated Sachs-Wolfe effect~\cite{Zentner:2001zr}.
In addition, too much relativistic energy affects the growth of large
scale structure (LSS) since the epoch for the matter-radiation equality
becomes later.  The increase of the total relativistic energy is
conventionally described by an effective number of light neutrino
species $\Delta N_\nu$.  The combined analysis of CMB and LSS data sets
the upper bound on $\Delta N_{\nu}$ as $\Delta N_\nu \le
5.0$~\cite{Ichikawa:2006vm}.  In Fig.~\ref{fig:cmb}, we show the
constraints from the CMB spectral distortion with $m_X =
10^3~\mathrm{GeV}$ when $B_X = 10^{-3}$ and $10^{-6}$.  In addition, we
also show the constraint from total relativistic energy injection, which
is independent of $B_X$.  When branching ratio is sufficiently small,
Fig.~\ref{fig:cmb} shows that the constraint from the total relativistic
energy injection provides severer constraint in a wide range of
lifetime.


\section{Diffuse neutrino and photon Constraints}
\label{sec:diffuse}

\subsection{Diffuse Neutrino Flux}
\label{subsec:neutrino_flux}

When neutrino injection takes place very late, 
the emitted neutrinos  may produce an observable peak 
in the diffuse neutrino spectrum.
The present differential flux of neutrinos is given 
by~\cite{Bergstrom:2001jj,Beacom:2006tt}
\begin{eqnarray}
  \frac{d\Phi_\nu}{dE_0}=\frac{1}{4\pi}
  \int dz\frac{1}{H_0h(z)}\frac{Y_Xs_0}{\tau_X}
  \exp(-t/\tau_X)\frac{dN_\nu}{dE_0},
  \label{neutrino_diffuse_flux}
\end{eqnarray}  
where $s_0$ is the present entropy density and
$h(z)=[(1+z)^3\Omega_m+\Omega_\Lambda]^{1/2}$
with $\Omega_m$ and $\Omega_\Lambda$ being the density
parameters of non-relativistic matter and dark energy, 
respectively.  Here neutrinos produced
with energy $E$ are redshifted to the observed energy $E_0=E/(1+z)$ and
$E$ is just half of the mass of $X$.  The source spectrum
$dN_\nu/dE_0$ is given by
\begin{eqnarray}
  \frac{dN_\nu(E_0)}{dE_0} =f_i\delta(E-E_0(1+z)),
  \label{neutrino_source}
\end{eqnarray}
where $f_i$ is the fraction of the neutrino spacies $i=e, \mu,
\tau$ emitted by the decay of $X$.  (Notice that $\sum_{i} f_i=1$.)  We
have assumed that only electron neutrinos are produced in the decay
process.  However, we should take the effects of neutrino oscillations
into account.  Neutrino oscillations can be described by six parameters:
two independent mass differences $( \Delta m^2_{12},\Delta m^2_{23})$,
three mixing angles $(\theta_{12},\theta_{23},\theta_{13})$ and a
CP-violating phase $\delta$.  A mixing angle $\theta_{23}$ is $\sim
45^\circ$ from atmospheric neutrino experiments~\cite{Ashie:2005ik}.
A mixing angle $\theta_{12}$ is determined by solar neutrino experiments
as $\theta_{12} \simeq 34^{\circ}$~\cite{Aharmim:2005gt}.  CHOOZ
experiment presented a mixing angle $\theta_{13} <
12^\circ$~\cite{Apollonio:2002gd}.  In our case, neutrino traveling
distance is very long and mass differences are irrelevant.
CP-violating phase $\delta$ enters the mixing matrix only in combination
with $\sin\theta_{13}$.  In a reasonable approximation, $f_e\sim0.6$ and
$f_\mu\sim f_\tau\sim0.2$.

\begin{figure}[thbp]
 \begin{center}
  \includegraphics[width=0.8\linewidth]{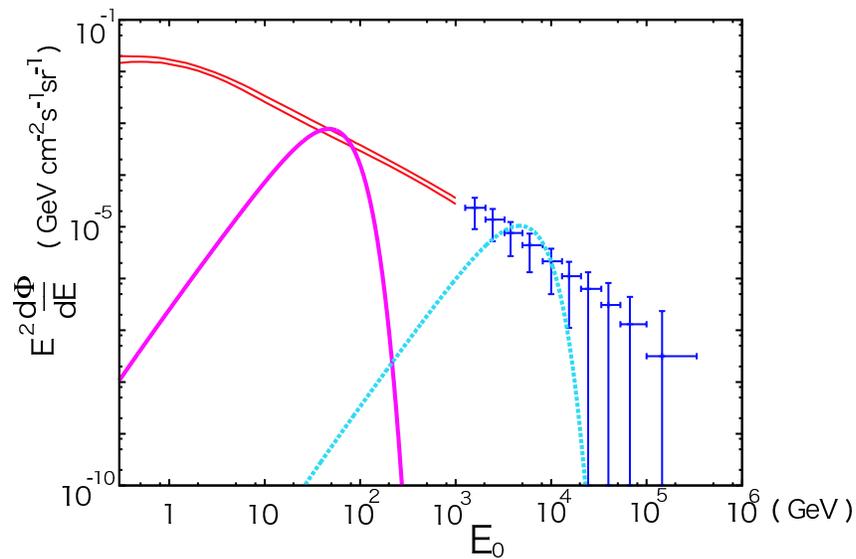}
  \vspace{-1cm}
                
  \caption{Atmospheric neutrino flux.  Thin solid lines represent $1
  \sigma$ range of the atmospheric neutrino fluxes
  \cite{Gonzalez-Garcia:2006ay}.  The point data are from AMANDA.  Thick
  solid and dotted lines represent diffuse neutrino signal with $m_X =
  10^3{\rm GeV}$ and $Y_X = 2.53\times10^{-17}$ (thick solid line) and
  $10^5{\rm GeV}$ and $Y_X = 3.37\times10^{-21}$(thick dotted line).
  The lifetime is $\tau_X=10^{16}{\rm sec}$.}  
  \label{fig:neutrino_flux}
 \end{center}
\end{figure}

The present
atmospheric neutrino $\nu_\mu+\bar{\nu}_\mu$ data gives the upper bound
of the differential flux of $\nu_\mu (\bar{\nu}_\mu)$ neutrinos.  The
atmospheric neutrino has been observed by
Super-Kamiokande~\cite{Ashie:2005ik} and AMANDA~\cite{Geenen:2003gg}.
For observational flux of energy range $0.3-1.0\times10^3~{\rm GeV}$ we
adopt the result in~\cite{Gonzalez-Garcia:2006ay} where the atmospheric
neutrino fluxes are estimated from the data on atmospheric neutrino
event rates measured by the Super-Kamiokande experiment.  For higher
ernergy range $1.3\times10^3-3.0\times10^5~{\rm GeV}$ we use the
atmospheric neutrino spectrum derived from AMANDA. In
Fig.~\ref{fig:neutrino_flux}, we show the atmospheric neutrino fluxes
from the Super-Kamiokande and AMANDA experiments as well as the diffuse
neutrino fluxes from the $X$ decay.

Futhermore, the diffuse neutrino flux is also constrained from null
detection of the relic supernova $\bar{\nu}_e$ flux by
Super-Kamiokande. In~\cite{Malek:2002ns}, the upperbound on
$\bar{\nu}_e$ flux is obtained as $\Phi_{\bar{\nu}_e}\le
1.2\mathrm{cm}^{-2}\mathrm{s}^{-1}$ above threshold of $ E_\nu >
19.3\mathrm{MeV}$.

\begin{figure}[thbp]
 \begin{center}
  \includegraphics[width=0.8\linewidth]{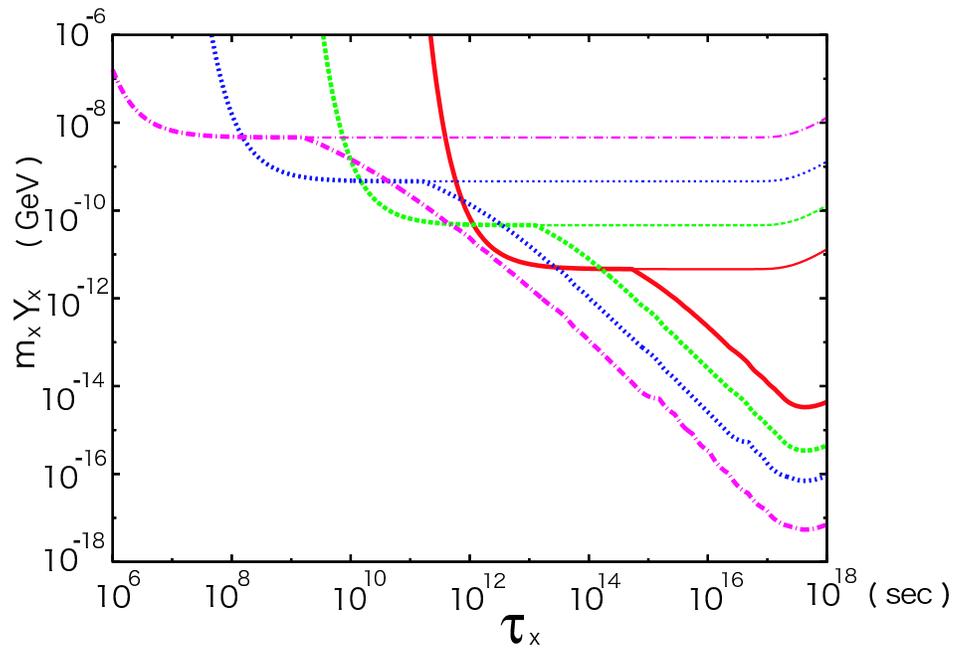}
  \caption{Constraints from diffuse neutrino flux.  From upper to lower
  the lines represent the upper bound of $X$ abundance when
  $m_X=10^2{\rm GeV}$ (solid line), $10^3{\rm GeV}$ (dashed line),
  $10^4{\rm GeV}$ (dotted line) and $10^5{\rm GeV}$ (dot-dashed line).
  Thin lines are constraints only from $\bar{\nu}_e$.}
  \label{fig:dif_neu}
 \end{center}
\end{figure}

We require that the neutrino flux from $X$ decay should not exceed the
observed atmospheric $\nu_\mu+\bar{\nu}_\mu$ flux and upperlimit of the
relic supernona $\bar{\nu}_e$ flux, which leads to the constraints on
the abundance of $X$ as shown in Fig.~\ref{fig:dif_neu}.  In the figure
we also show the constraint from relic supernova search only.  When
lifetime is short, the constraint is very weak since the neutrinos get
redshifted until the present and their energy becomes lower than the
$20$~MeV. For intermediate lifetime, the constraints on the abundances
of $X$ are determined by the relic supernova $\bar{\nu}_e$ search and
they are in proportion to $m_X$ because the diffuse $\bar{\nu}_e$ flux
is determined by the number of injected neutrinos above threshold
energy, hence $Y_X$.  When lifetime is long, however, the constraints
are determined by atmospheric neutrino fluxes.  Since the neutrino flux
has a peak at $\sim m_X /(1+z_{d})$ ($z_d$: redshift at $t=\tau_X$), the
maximum differential neutrino flux is propotional to $Y_X/m_X$. On the
other hand, the observed differential neutrino flux is roughly
propotional to $E^{-3}$ and hence $m_X^{-3}$ at $E\sim m_X$. Therefore,
the constraint on $m_X Y_X$ depends on $m_X$ as $m_X m_X^{-3}/m_X^{-1}
\sim m_X^{-1}$.
This means the constraints become severe with larger mass as shown in
Fig.~\ref{fig:dif_neu}.  Constraints for lifetimes longer than present
time $t_0$ scale by a factor $t_0/\tau_X$ relative to the constraints at
$t_0=\tau_X$.


\subsection{Diffuse Photon Flux}

High energy photons and electrons (positrons) produced in the three and
four-body decay may be observed as diffuse gamma rays when the decay
takes place after the recombination epoch.  In calculating the diffuse
photon spectrum, we must consider primary photon spectrum which is not
monoenergetic.  In addition, we should also take account of the
absorption of gamma rays along the line of sight.  

\begin{figure}[thbp]
  \begin{center}
   \includegraphics[width=0.8\linewidth]{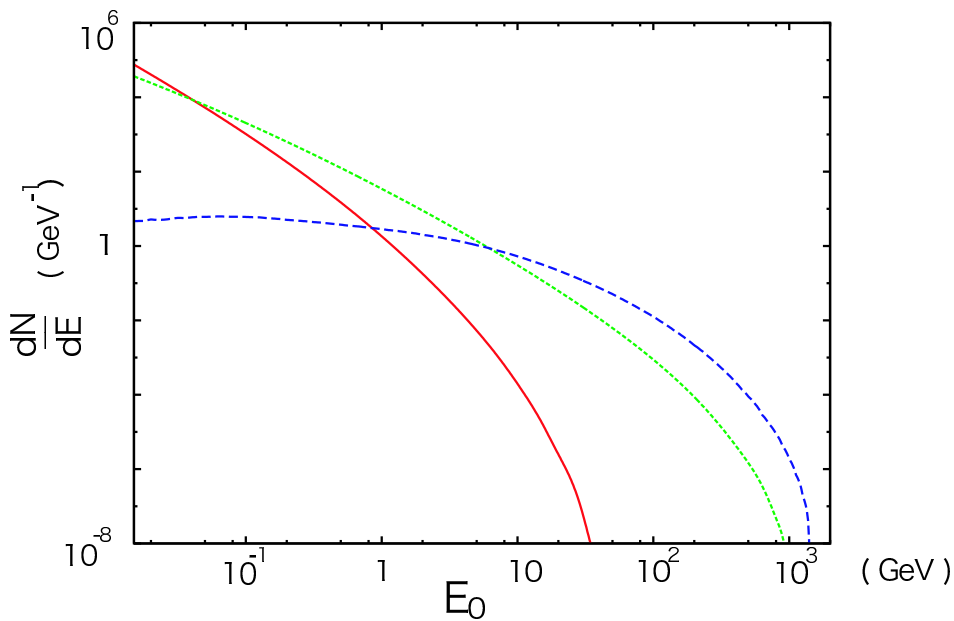} \caption{Photon
   spectrum versus photon energy for $m_X=10^4{\rm GeV}$.  Solid line
   represents photon produced through inverse Compton process at $1+z=1$
   and dotted line represents at $1+z=100$.  Dashed line represents
   photon produced by direct decay of $X$.}  \label{fig:phot_spec}
  \end{center}
\end{figure}

Since photons are produced through three- and four-body decays, their
spectrum is not monochromatic unlike neutrino.  The energy distributions
of photons, neutrinos, leptons and nucleons produced by the three- and
four-body decay of $X$ are shown in
Figs.~\ref{fig:decay_m=1e2}-\ref{fig:decay_m=1e5}.  The energy of
produced electrons and positrons is transferred to the background
photons through inverse Compton process.
In calculating the photon flux, we have taken into account the photons
produced by the inverse Compton process as well as those from the
cascade decay chain induced by the three- and four-body decay of $X$.
The latter effect becomes more important for high energy photons.
However, since the inverse Compton process produces many soft photons,
the formar process becomes more significant for low energy photons.  The
details of calculation of inverse Compton process are given
in~\ref{app:inv}.  Nucleons are also produce in the decay and they
produce photons through inverse Compton process or $\beta$
decay. However, we neglect this effect since the number density of
produced nucleons is sufficiently small.  In Fig.~\ref{fig:phot_spec},
we plot photon flux for $m_X=10^4{\rm GeV}$.


High energy photons
injected in the universe, in general, scatter through various processes;
photon pair production, photon-photon scattering and pair production in
matter.  At the eraly epochs absorption and scattering due to the
background photons are important, whereas significant absorption by
diffuse IR-UV photons emitted from galaxies takes place at
later epochs.

First let us consider the radiative processes due to the background
photons.  In the present case, the relevant processes are photon-photon
scattering and photon pair creation~\cite{1989ApJ...344..551Z}.  Then,
photons degrade their energy by producing electron-positron pairs or
dividing their energy with the background photons.  The photon spectrum
was calculated in detail in~\cite{Kribs:1996ac}. According
to~\cite{Kribs:1996ac} the present differential flux of photons is given
by
\begin{eqnarray}
        \frac{d\Phi_\gamma}{dE_0} &= &
        \frac{1}{4\pi}\int^{z_*}_0 dz\frac{1}{H_0h(z)}
        \frac{Y_Xs_0}{\tau_X}\exp(-t/\tau_X) \nonumber \\
         & & \times \int dE \mathcal{L}_s(E_0(1+z),E,z)
         \mathcal{L}_i(E,z)B_X,
         \label{eq:photon_flux}
\end{eqnarray}
where $\mathcal{L}_i(E,z)$ is the number of photons per unit energy
produced both directly and by inverse Compton at $z$ for one $X$ decay,
and $z_*( \simeq 700)$ is the redshift at which the optical depth of
high enrergy photons becomes $1$.  In addition,
$\mathcal{L}_s(E_1,E_2,z)$ is the number of photons per unit energy in
the spectrum when photons with energy $E_1$ are produced by scattering
of photons with energy $E_2$ at redshift $z$.  If there is no
scattering, $\mathcal{L}_s(E_1,E_2,z)$ becomes $\delta(E_1-E_2)$. (The
concrete expression of $\mathcal{L}_s(E_1,E_2,z)$ is found in
\cite{Kribs:1996ac}.)

In addition, $\gamma$ rays with GeV to TeV energies are absorbed via
electron-positron pair production on diffuse background IR-UV photons
which have been emitted by
galaxies~\cite{Salamon:1997ac,Primack:2000xp,Stecker:2005qs}.  In this
paper, we adopt the result of \cite{Stecker:2005qs} which calculated the
optical depth $\tau_{\rm IR}$ of the universe for $\gamma$ rays having
energies from 4 GeV to 100 TeV at redshifts from 0 to 5.  (See Fig.~8 in
\cite{Stecker:2005qs}.)  Then the resultant photon spectrum is given by
Eq.~(\ref{eq:photon_flux}) multiplied by $e^{-\tau_{\rm IR}}$.  We
neglect the secondary soft photons produced via
electron-positron pair production on diffuse background IR-UV photons,
and only consider the attenuation of high energy photons.

\begin{figure}[thbp]
 \begin{center}
  \includegraphics[width=0.8\linewidth]{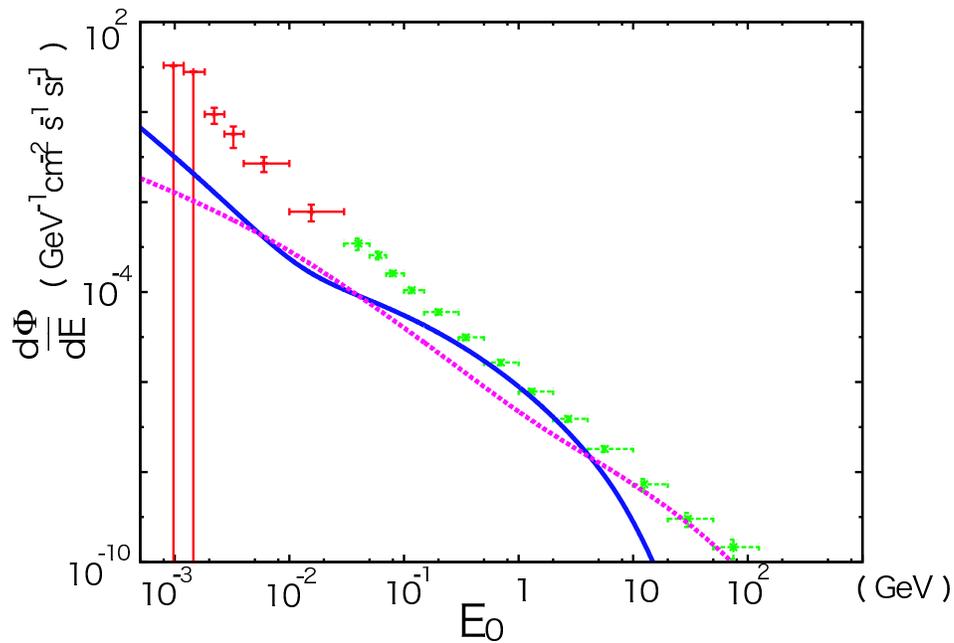}
  \caption{Diffuse photon flux. The point data are from 
  COMPTEL and EGRET.
  Solid and dotted lines represent diffuse photon signal 
  for $m_X = 10^3~{\rm GeV}$ and $B_XY_X = 8.86\times10^{-20}$(solid line) 
  and $10^5~{\rm GeV}$ and $B_XY_X = 2.00\times10^{-21}$(dotted line). 
  The lifetime is $\tau_X=10^{16}~{\rm sec}$.}
  \label{fig:photon_flux}
 \end{center}
\end{figure}
\begin{figure}[thbp]
 \begin{center}
  \includegraphics[width=0.8\linewidth]{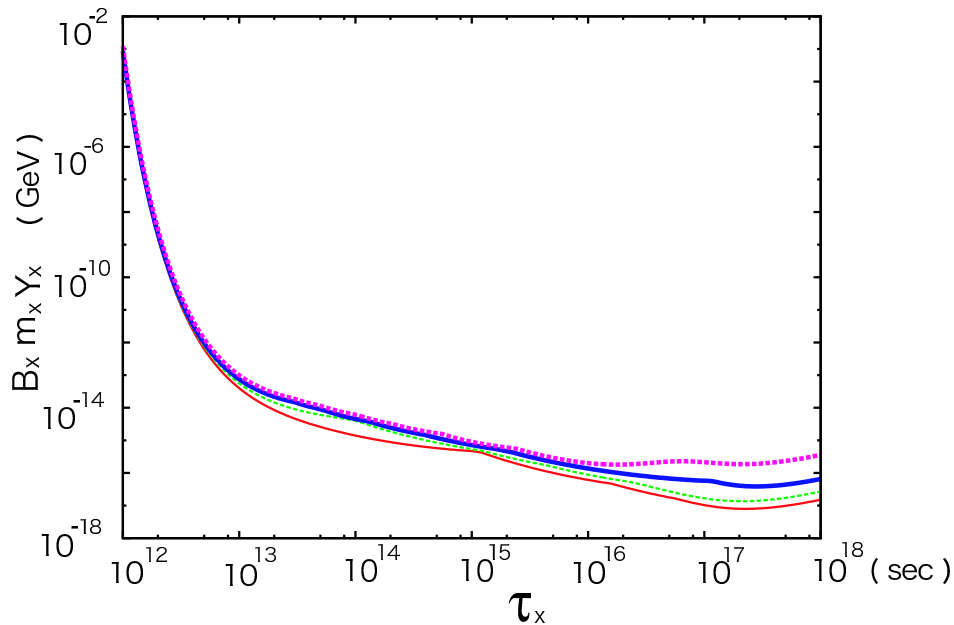}
  \caption{Constraints from diffuse photon flux. 
  From upper to lower the lines represent 
  the upper bound of $X$ abundance when $m_X=10^2~{\rm GeV}$ 
  (thin solid line), $10^3~{\rm GeV}$ (thin dotted line), 
  $10^4~{\rm GeV}$ (thick solid line) 
  and $10^5~{\rm GeV}$ (thick dotted line).}
  \label{fig:dif_phot}
 \end{center}
\end{figure}

In Fig.~\ref{fig:photon_flux}, we show the diffuse photon flux from the
COMPTEL \cite{1996A&AS..120C.619K} and EGRET \cite{1998ApJ...494..523S}
observations and decay of $X$, from which we obtain the upper limit on
the abundance of $X$ as shown in Fig.~\ref{fig:dif_phot}.  Compared with
the limit from the neutrino flux, the constraints from photon flux are
almost insensitive to $m_X$. This can be understood as follows.  The
differential photon flux at the peak energy is roughly proportional to
$Y_X/m_X$ from the same reasoning as the neutrino flux in
Sec.~\ref{subsec:neutrino_flux}, while the observed one is proportional
to $E^{-2} \propto m_X^{-2}$. Thus, the constraint on $Y_X$ depends on $m_X$
as $m_X^{-2}m_X \sim m_X^{-1}$, which means that the limit on $m_X Y_X$
is almost insensitive to $m_X$. 
Fig.~\ref{fig:dif_phot} shows that the constraints become less stringent
with larger $m_X$ when lifetime is long.  There are two reasons for this.
One is EGRET had observed up to $100{\rm GeV}$.  When $m_X$ is large,
the present photon energy at which the flux becomes maximum  may exceed
the energy range of EGRET observation.  The other is higher energy photons are
more effectively absorbed by diffuse background photons.  


\section{Conclusion}
\label{sec:conclusion}

\begin{figure}[thbp]
 \begin{center}
  \includegraphics[width=0.8\linewidth]{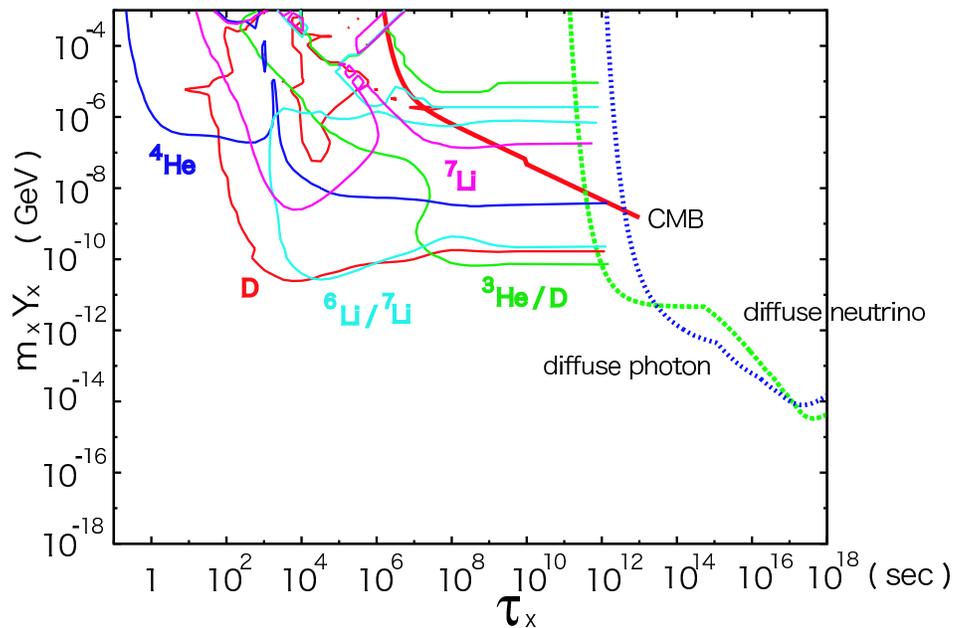} 
  \caption{The constraints on the relic abundance of $X$ from 
  various observations with
  $m_X=100{\rm GeV}$ and $B_X=10^{-3}$.}  \label{fig:Const_all_E2B3}
 \end{center}
\end{figure}

\begin{figure}[thbp]
 \begin{center}
  \includegraphics[width=0.8\linewidth]{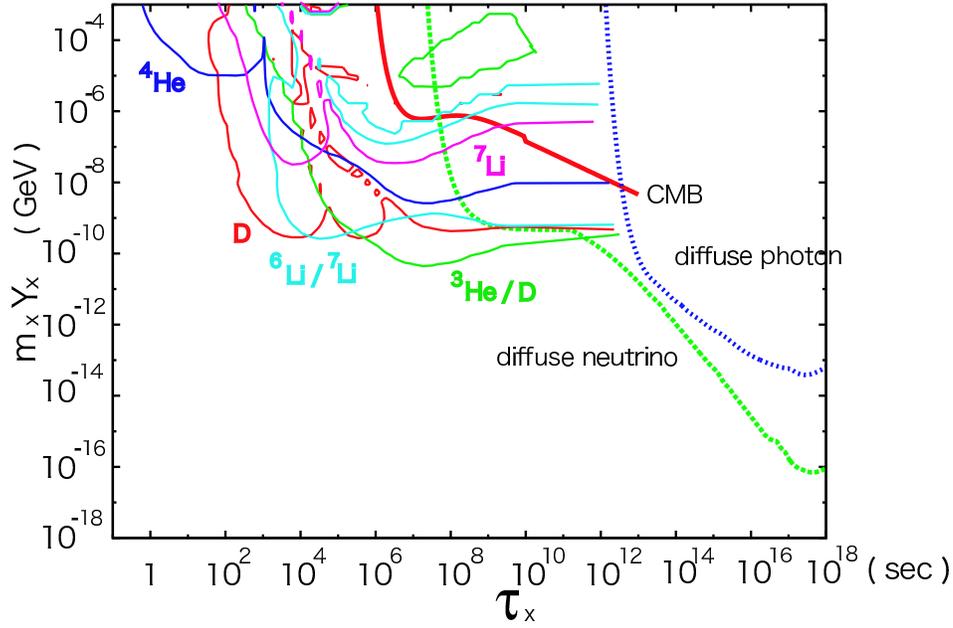}
  \caption{The constraints on the relic abundance of $X$ from 
  various observations 
  with $m_X=10{\rm TeV}$ and $B_X=10^{-3}$.}
  \label{fig:Const_all_E4B3}
 \end{center}
\end{figure}

In this paper, we have considered the long-lived massive particle $X$
which mainly decays into a neutrino and an invisible particle, and have
investigated the cosmological and astrophysical constraints on the
high-energy neutrino and photon injection due to decay of $X$-particle.
We have shown that the BBN, CMB, diffuse neutrino fluxes and diffuse
gamma rays provide stringent constraints on the abundance of the
decaying particle $X$.  We summarize the constraints in
Figs.~\ref{fig:Const_all_E2B3} and ~\ref{fig:Const_all_E4B3}.  


\vspace{1cm}

\noindent
{\it Acknowledgments}: This work was supported in part by the
Grant-in-Aid for Scientific Research from the Ministry of Education,
Science, Sports, and Culture of Japan, No.\ 18540254 and No 14102004
(M.K.).  This work was also supported in
part by JSPS-AF Japan-Finland Bilateral Core Program (M.K.), and PPARC
grant, PP/D000394/1, EU grant MRTN-CT-2006-035863, the European Union
through the Marie Curie Research and Training Network "UniverseNet",
MRTN-CT-2006-035863 (K.K.).

\appendix

\section{Bottzmann equation}
\label{app:Boltzmann}

In order to investigate effects of photo-dissociation processes and
$p\leftrightarrow n$ conversion processes, we have to calculate a photon
spectrum.  In this paper, a source of high energy photon is charged
leptons and pions which are produced through scattering of high energy
neutrinos off background leptons.  Therefore, we determine the time
evolution of the distribution function of high energy neutrinos in order
to investigate the photon spectrum.  In this appendix, we write down
Boltzmann equations which determines the high energy neutrino spectrum.
Our notation is the same as~\cite{Moroi:1995fs}.

The high energy neutrinos ($\nu$) produced in $X$ decay scatter off
the thermal neutrino ($\nu_b$) in the background by the following processes;
\begin{eqnarray}
    \nu_{i} + \nu_{i,b} &\rightarrow & \nu_i + \nu_i , 
     \label{reaction:neu-neu1} \\
    \nu_{i} + \bar{\nu}_{i,b} &\rightarrow &\nu_i + \bar{\nu}_i,\\
    \nu_{i} + \bar{\nu}_{i,b} &\rightarrow &\nu_j + \bar{\nu}_j ,\\
    \nu_{i} + \nu_{j,b} &\rightarrow &\nu_i + \nu_j,\\
    \nu_{i} + \bar{\nu}_{j,b} & \rightarrow &\nu_i + \bar{\nu}_j, 
     \label{reaction:neu-neu2} \\
    \nu_{i} + \bar{\nu}_{i,b} &\rightarrow & e^{-}  + e^{+},\\
    \nu_{i} + \bar{\nu}_{i,b} & \rightarrow & \mu^{-}  + \mu^{+}.
\end{eqnarray}
where index $i$ and $j$ represent $e$, $\mu$ and $\tau$ with $i\neq j$.
All the amplitude squared $|{\cal M}|^2$ in these reactions take the
following form:
\begin{eqnarray}
    |{\cal M}|^2 = 32G_F^2[ 
    a(pp')^2 + b(pq)^2 + c(pq')^2 + d m^2 (pp')] ,
    \label{amplitude}
\end{eqnarray}
where $G_F \simeq 1.17\times 10^{-5}{\rm GeV}^{-2}$ is the Fermi constant,
the coefficients $a$ -- $d$ depend on the individual reaction, $p$ and
$p'$ are the initial momenta of high energy neutrino and background
neutrino, $q$ and $q'$ are the final momenta, and $m$ represents the
mass of the fermion the in final state. 
Coefficients for each processes are given in Table~\ref{table:neutrino-neutrino}. 
%
%
\begin{table}
\begin{center}
\begin{tabular}{c|lccc} \hline \hline
{Process} &
{~~a~~~~~~~~~} &
{~~~~~~~~~b~~~~~~~~~} &
{~~~~~~~~~c~~~~~~~~~} &
{~~~~~~~~~d~~~~~~~~~} \\ \hline
{$\nu_{i} + \nu_{i,b} \rightarrow \nu_i + \nu_i$} &
{~~2} &  {0} & {0} &  {0} \\
{$\nu_{i} + \bar{\nu}_{i,b} \rightarrow \nu_i + \bar{\nu}_i$} &
{~~0} &  {0} & {9} &  {0} \\
{$\nu_{i} + \bar{\nu}_{i,b} \rightarrow \nu_j + \bar{\nu}_j$} &
{~~0} &  {0} & {1} &  {0} \\
{$\nu_{i} + \nu_{j,b} \rightarrow \nu_i + \nu_j$} &
{~~1} &  {0} & {0} &  {0} \\
{$\nu_{i} + \bar{\nu}_{j,b} \rightarrow \nu_i + \bar{\nu}_j$} &
{~~0} &  {0} & {1} &  {0} \\
{$\nu_{i} + \bar{\nu}_{i,b} \rightarrow l_i^{-}  + l_i^{+}$} &
{~~0} & {$(C_V-C_A)^2$} & {$(C_V+C_A+2)^2$} & {$(C_V-C_A)(C_V+C_A+2)$} \\
{$\nu_{i} + \bar{\nu}_{i,b} \rightarrow l_j^{-}  + l_j^{+}$} &
{~~0} & {$(C_V-C_A)^2$} & {$(C_V+C_A)^2$} & {$C^2_V-C^2_A$} \\
\hline \hline
\end{tabular}
\caption{Coefficients $a$ -- $d$ for each processes. Index $i$ and $j$
(with $i\neq j$) represent the generation, $\nu_{i,b}$ is the background
neutrino of $i$-th generation, $l^{\pm}_i$ is the charged lepton of
$i$-th generation (in our case, $e^{\pm}$ or $\mu^{\pm}$).  $C_V$ and
$C_A$ are defined as follows: $C_V=-0.5+2\sin\theta^2_W$ and $C_A=-0.5$.
Here, $\theta_W$ is the Weinberg angle.}
\label{table:neutrino-neutrino}
\end{center}
\end{table}
%
%

First, let us consider the neutrino scattering processes
Eqs.~(\ref{reaction:neu-neu1})-(\ref{reaction:neu-neu2}).  Here, we
define $E_\nu$ as the energy of initial high energy neutrino and
$E'_\nu$ as the energy of neutrino in final state.  We also write the
distribution function of the background neutrino as
\begin{eqnarray}
 \bar{f}_\nu(\bar{E}_\nu)=
  \frac{\bar{E}^2_\nu}{2\pi^2}\frac{1}{\exp(\bar{E}_\nu/T_\nu)+1},
\end{eqnarray}  
where $\bar{E}_\nu$ is the energy of background neutrino and $T_\nu$ is
the neutrino temperature.  We describe the contribution to the time
derivative of the neutrino distribution function.  When
$E_\nu,E'_\nu \gg \bar{E}_\nu$, the increases of distribution function
due to scattering is written as
\begin{eqnarray}
 \left.\frac{\partial f_\nu(E'_\nu)}{\partial t}\right|_{+}
  &=&\frac{4}{3\pi}G^2_F
  \int^{\infty}_{E'_\nu}dE_\nu\frac{1}{E^2_\nu}
  [aE^2_\nu+b(E_\nu-E'_\nu)^2+c{E'}^2_\nu]f_\nu(E_\nu)
  \nonumber \\
  &\times& \int^{\infty}_0 d\bar{E}_\nu \bar{E}_\nu 
  \bar{f}_\nu(\bar{E}_\nu)
\end{eqnarray}  
On the other hand, decreases of neutrino distribution function is written
as
\begin{eqnarray}
 \left.\frac{\partial f_\nu(E_\nu)}{\partial t}\right|_{-}
  &=&-\frac{1}{8}\frac{1}{E^2_\nu}f_\nu(E_\nu)
  \int^{\infty}_{0}d\bar{E}_\nu\frac{1}{\bar{E}^2_\nu}
  \bar{f}_\nu(\bar{E}_\nu)
  \int^{4E_\nu\bar{E}_\nu}_{0}ds s\sigma(s) 
  \nonumber \\
 &=& -\frac{4}{3\pi}G^2_F
 \left(a+\frac{1}{3}b+\frac{1}{3}c\right)
 \int^{\infty}_0 d\bar{E}_\nu \bar{E}_\nu \bar{f}_\nu(\bar{E}_\nu),
\end{eqnarray}  
where $\sigma(s)$ is the total cross section obtained from the amplitude
Eq.~(\ref{amplitude}).  Notice that the condition for the neutrino
number conservation is realized;
\begin{eqnarray}
 \int^{\infty}_0 dE_\nu
  \left.\frac{\partial f_\nu(E_\nu)}{\partial t}\right|_{+}
  = -\int^{\infty}_0 dE_\nu
  \left.\frac{\partial f_\nu(E_\nu)}{\partial t}\right|_{-}
\end{eqnarray}
unless the effects of inelastic channels $(\nu+\bar{\nu}\to
e^{+}+e^{-},\mu^{+}+\mu^{-})$ are taken into account.  

Effects of thecharged lepton pair creation process can be taken into account in the same way, 
and the contribution to the time derivative of the neutrino
distribution function is given by
\begin{eqnarray}
 \left.\frac{\partial f_\nu(E_\nu)}{\partial t}\right|_{-}
  &=&-\frac{1}{8}\frac{1}{E^2_\nu}f_\nu(E_\nu)
  \int^{\infty}_{0}d\bar{E}_\nu\frac{1}{\bar{E}^2_\nu}
  \bar{f}_\nu(\bar{E}_\nu)
  \int^{4E_\nu\bar{E}_\nu}_{4m^2}ds s\sigma(s) 
  \nonumber \\
 &=& -\frac{1}{16\pi}G^2_F
 \frac{1}{E^2_\nu}f_\nu(E_\nu)
 \int^{\infty}_0 d\bar{E}_\nu \frac{1}{\bar{E}^2_\nu} 
 \bar{f}_\nu(\bar{E}_\nu)
 \nonumber \\
 &\times&\left[\left(a+\frac{1}{3}b+\frac{1}{3}c\right)I_2
          +\left(2d-\frac{1}{3}b-\frac{1}{3}c\right)m^2I_1\right]
 \label{pair-creation}  
\end{eqnarray}  
with
\begin{eqnarray}
 I_2&=&\frac{4}{3}
  \left(4-\frac{4m^2}{E_\nu\bar{E}_\nu}\right)^{1/2}E_\nu\bar{E}_\nu
  (8E^2_\nu\bar{E}^2_\nu-2m^2E_\nu\bar{E}_\nu-3m^4)
  \nonumber \\
 &-&4m^6\ln\left[\frac{2\{4-(4m^2/E_\nu\bar{E}_\nu)\}^{1/2}E_\nu\bar{E}_\nu
            +4E_\nu\bar{E}_\nu-2m^2}{2m^2}\right],
\\
 I_1&=&2\left(4-\frac{4m^2}{E_\nu\bar{E}_\nu}\right)^{1/2}E_\nu\bar{E}_\nu
 (2E_\nu\bar{E}_\nu-m^2)
 \nonumber \\
 &-&2m^4\ln\left[\frac{2\{4-(4m^2/E_\nu\bar{E}_\nu)\}^{1/2}E_\nu\bar{E}_\nu
            +4E_\nu\bar{E}_\nu-2m^2}{2m^2}\right].      
\end{eqnarray}
Coefficients for the charged lepton production processes are given 
in Table~\ref{table:neutrino-neutrino}. 

In addition, there are processes for neutrino scattering as follows:
\begin{eqnarray}
 \nu_{e}+\bar{\nu}_{\mu,b}\to\mu^{+}+e^{-}, \\
 \nu_{\mu}+\bar{\nu}_{e,b}\to\mu^{-}+e^{+}.
\end{eqnarray}  
The amplitudes squared in these reactions take the form given by
\begin{eqnarray}
 |{\cal M}|^2_{\nu_{e}+\bar{\nu}_{\mu,b}\to\mu^{+}+e^{-}} 
  &=& 128G_F^2(pq')
  \left[(pq')-\frac{1}{2}(m^2_\mu-m^2_e)\right],\\
 |{\cal M}|^2_{\nu_{\mu}+\bar{\nu}_{e,b}\to\mu^{-}+e^{+}} 
  &=& 128G_F^2(pq')
  \left[(pq')+\frac{1}{2}(m^2_\mu-m^2_e)\right].
\end{eqnarray}          
Effects of these process can be taken into account in the same way.
However, it is somewhat complicated to calculate Boltzmann equations for
these processes due to the mass difference between muon and electron.
The contributions to the time derivative of the neutrino distribution
function are given by
\begin{eqnarray}
 \left.\frac{f_{\nu_e}(E_\nu)}{\partial t}\right|_{\nu_{e}
  +\bar{\nu}_{\mu,b}\to\mu^{+}+e^{-}}
  &=&
  \left.\frac{f_{\nu_\mu}(E_\nu)}{\partial t}
   \right|_{\nu_{\mu}+\bar{\nu}_{e,b}\to\mu^{-}+e^{+}},
   \nonumber  \\ 
 &=& -\frac{G_F^2}{24\pi}E_\nu f_{\nu}(E_\nu)
 \int^{\infty}_0 d \bar{E}_\nu \bar{f}_\nu (\bar{E}_\nu)I_3.
\end{eqnarray}
with
\begin{eqnarray}
 I_3 &=& \frac{z}{3}\left[-5x^2 - 2xy -5y^2 - 20(x + y) + 64\right]
  \nonumber \\ 
 &+& \left(x^3 - 3x^2y -3xy^2 +y^3\right)\ln
 \left[\frac{4 - x - y + 2z}{2\sqrt{xy}}\right]
 \nonumber \\ 
 &+&|x - y|^3\ln\left[\frac{4(x + y)-(x - y)^2-2z|x-y|}{8\sqrt{xy}}\right].
\end{eqnarray}
where $x = m^2_\mu/E_\nu\bar{E}_\nu$, $y = m^2_e/E_\nu\bar{E}_\nu$ and 
$z = \left[4 - 2(x + y) + \frac{1}{4}(x - y)^2\right]^{\frac{1}{2}}$. 

Finally, we consider the effect of charged pion pair creation process.
The cross section for charged pion pair production is given by
\begin{eqnarray}
 \sigma(\nu\bar{\nu}\to\pi^{+}\pi^{-})
  =\frac{1}{12\pi}G_F^2(1-2\sin^2\theta_W)^2 
  s\left(1 - \frac{4m^2_\pi}{s}\right)^{\frac{3}{2}}|F(s)|^2.
\end{eqnarray}
with
\begin{eqnarray}
 |F(s)|^2=\frac{m^4_\rho}{(s - m^2_\rho)^2 + m^2_\rho\Gamma^2_\rho}.
  \label{definition:F}
\end{eqnarray}
where $m_\rho$ and $\Gamma_\rho$ is the mass and decay width of the
$\rho$ meson, respectively~\cite{Feynman:1972}.  The contribution to the
time derivative of the neutrino distribution function is given by
Eq.~(\ref{pair-creation}) and the cross section above.

Next, we turn now to neutrino-electron scattering processes.  The high
energy neutrinos also scatter off the thermal electron $(e^{-}_b)$ and
positron $(e^{+}_b)$ by the following processes;
\begin{eqnarray}
 \nu_{e} + e^{-}_{b} &\rightarrow & \nu_{e} + e^{-} , 
  \label{reaction:neu-elec1} \\
 \nu_{e} + e^{+}_{b} &\rightarrow & \nu_{e} + e^{+},\\
 \nu_{i} + e^{-}_{b} &\rightarrow &  \nu_{i} + e^{-} ,\\
 \nu_{i} + e^{+}_{b} &\rightarrow & \nu_{i} + e^{+}. 
  \label{reaction:neu-elec2}
\end{eqnarray}
where index $i$ represents $\mu$ and $\tau$.  All the amplitude squared
in these reactions take the same form as Eq.~(\ref{amplitude}).  In this
case, however, $p$ and $q$ are the initial and final momenta of neutrino
and $p'$ and $q'$ are that of background electron (positron).
Coefficients for each processes are given in
Table~\ref{table:neutrino-electron}.
%
%
\begin{table}
\begin{center}
\begin{tabular}{c|lccc} \hline \hline
{Process} &
{~~~~~~~~~a~~~~~~~~~} &
{~~~~~~b~~~~~} &
{~~~~~~~~~c~~~~~~~~~} &
{~~~~~~~~~d~~~~~~~~~} \\ \hline
{$\nu_{e} + e^{-}_{b} \rightarrow  \nu_{e} + e^{-}$} &
{$(C_V+C_A+2)^2$} &  {0} & {$(C_V-C_A)^2$} &  {$-(C_V-C_A)(C_V+C_A+2)$} \\
{$\nu_{e} + e^{+}_{b} \rightarrow  \nu_{e} + e^{+}$} &
{$(C_V-C_A)^2$} &  {0} & {$(C_V+C_A+2)^2$} &  {$-(C_V-C_A)(C_V+C_A+2)$} \\
{$\nu_{i} + e^{-}_{b} \rightarrow   \nu_{i} + e^{-}$} &
{$(C_V+C_A)^2$} &  {0} & {$(C_V-C_A)^2$} &  {$-(C^2_V-C^2_A)$} \\
{$\nu_{i} + e^{+}_{b} \rightarrow  \nu_{i} + e^{+}$} &
{$(C_V-C_A)^2$} &  {0} & {$(C_V+C_A)^2$} &  {$-(C^2_V-C^2_A)$} \\
\hline \hline
\end{tabular}
\caption{Coefficients $a$ -- $d$ for each processes. Index $i$ 
represents $\mu$ and $\tau$, $e_{b}$ is the background
electron and positron.}
\label{table:neutrino-electron}
\end{center}
\end{table}
%
%
We write the distribution function of the background electron (positron)
as
\begin{eqnarray}
 \bar{f}_e(\bar{E}_e)=\frac{\bar{p}^2_e}{2\pi^2}
  \frac{1}{\exp(\bar{E}_e/T_\gamma)+1}.
\end{eqnarray}
where $\bar{E}_e$ and $\bar{p}_e$ is the energy and momentum of
background electron (positron), respectively.  Photon temperature
$T_\gamma$ is different from neutrino temperature $T_\nu$ due to
neutrino decoupling and subsequent electron-positron pair annihilation.
From entropy conservation, the relation between them is given by
\begin{eqnarray}
 T_\nu=\left(\frac{4}{11}\right)^{1/3}T_\gamma
  \left[1+\frac{45}{2\pi^2}\frac{1}{T^4_\gamma}
   \int^{\infty}_0 d\bar{p}_e
   \left(\bar{E}_e+\frac{\bar{p}^2_e}{3\bar{E}_e}\right)
   \bar{f}_e(\bar{E}_e)\right]^{1/3}.
\end{eqnarray}
The effects of neutrino-electron scattering are only important at early
time because of the Boltzmann suppression of the distribution function
of background electron (positron).  When $E_\nu,E'_\nu\gg\bar{E}_e$, the
increase and decrease of distribution functions are given by
\begin{eqnarray}
 \left.\frac{\partial f_\nu(E'_\nu)}{\partial t}\right|_{+}
  &=&\frac{4}{3\pi}G^2_F
  \int^{\infty}_{E'_\nu}dE_\nu\frac{1}{E^2_\nu}
  [aE^2_\nu+b(E_\nu-E'_\nu)^2+c{E'}^2_\nu]f_\nu(E_\nu)
  \nonumber \\
 &\times& \int^{\infty}_0 d\bar{p}_e \bar{E}_e \bar{f}_e(\bar{E}_e)
 \left(1-\frac{m^2_e}{4\bar{E}^2_e}\right),
\\
 \left.\frac{\partial f_\nu(E_\nu)}{\partial t}\right|_{-}
 &=&-\frac{1}{8}\frac{1}{E^2_\nu}
 f_\nu(E_\nu)\int^{\infty}_0 d\bar{p}_e
 \frac{1}{\bar{p}_e\bar{E}_e}\bar{f}_e(\bar{E}_e)
 \int^{} ds (s-m^2_e)\sigma(s)
 \nonumber \\
 &=&-\frac{4}{3\pi}G^2_F
 \left(a+\frac{1}{3}b+\frac{1}{3}c\right)E_\nu f_\nu(E_\nu)
 \nonumber \\
 &\times& \int^{\infty}_0 d\bar{p}_e \bar{E}_e \bar{f}_e(\bar{E}_e)
 \left(1-\frac{m^2_e}{4\bar{E}^2_e}\right).
 \label{pair-creation2}         
\end{eqnarray}  
Notice that the condition for the neutrino number conservation is also
realized just as neutrino-neutrino scattering.

We also include the following process:
\begin{eqnarray}
 \nu_\mu + e^{-}\to \nu_e+\mu^{-}.
\end{eqnarray}
The amplitude squared in this reaction is give by
\begin{eqnarray}
 |{\cal M}|^2_{\nu_\mu + e^{-}\to \nu_e+\mu^{-}} &=& 128G_F^2(pp')
  \left[(pp')-\frac{1}{2}(m^2_\mu-m^2_e)\right].
\end{eqnarray}          
For simplicity, we neglect the mass difference between muon and
electron in this reaction.  On this assumption, this process is the same
form as $\nu_i+e\to\nu_i+e$.

In addition, we consider the effect of pion pair creation process.
The cross section for pion pair production is given by
\begin{eqnarray}
 \sigma(\bar{\nu}_e e^{-}\to\pi^{-}\pi^{0})=\frac{1}{12\pi}G_F^2
  s\left(1 - \frac{4m^2_\pi}{s}\right)^{\frac{3}{2}}|F(s)|^2,
\end{eqnarray}
where $|F(s)|^2$ is defined in Eq.~(\ref{definition:F})~\cite{Feynman:1972}.
The contribution to the time derivative of the neutrino distribution function 
is given by Eq.~(\ref{pair-creation2}) with the cross section above.    

Then one can obtain the Boltzmann equations describing the evolution of
the spectra for the high energy neutrinos;
\begin{eqnarray}
 \frac{\partial f_{\nu_i}(E'_\nu)}{\partial t} &=& 
  \frac{4G_F^2}{3\pi}
  \int^{\infty}_{E'_\nu} dE_\nu\frac{1}{E^2_\nu}
  \sum_j\left[a^{\nu}_{in,ij}E^2_\nu+b^{\nu}_{in,ij}(E_\nu-E'_\nu)^2 
         +c^{\nu}_{in,ij}{E'}^2_\nu\right]f_{\nu_j}(E_\nu)
  \nonumber \\
 &\times& \int^{\infty}_0 d\bar{E}_{\nu} 
 \bar{E}_{\nu}\bar{f}_{\nu}(\bar{E}_{\nu}) 
 \nonumber \\
 &-&\frac{4G_F^2}{3\pi} E'_\nu f_{\nu_i}(E'_\nu) 
 \left(a^\nu_{out}+\frac{1}{3}b^\nu_{out}+\frac{1}{3}c^\nu_{out}\right)
 \int^{\infty}_0 d\bar{E}_{\nu} \bar{E}_{\nu}\bar{f}_{\nu}(\bar{E}_{\nu})
\nonumber \\
 &+&\frac{4G_F^2}{3\pi}
 \int^{\infty}_{E'_\nu} dE_\nu\frac{1}{E^2_\nu}
 \left[a^{e}_{in,i}E^2_\nu+b^{e}_{in,i}(E_\nu-E'_\nu)^2 
  +c^{e}_{in,i}{E'}^2_\nu\right]f_{\nu_i}(E_\nu)
 \nonumber \\
 &\times& \int^{\infty}_0 d\bar{p}_e \bar{E}_e 
 \bar{f}_e(\bar{E}_e)\left(1-\frac{1}{4}\frac{m^2_e}{\bar{E}^2_e }\right) 
 \nonumber \\
 &-&\frac{4G_F^2}{3\pi} E'_\nu f_{\nu_i}(E'_\nu) 
 \left(a^{e}_{out,i}+\frac{1}{3}b^{e}_{out,i}+\frac{1}{3}c^{e}_{out,i}\right)
 \int^{\infty}_0 d\bar{p}_{e} \bar{E}_{e}\bar{f}_{e}(\bar{E}_{e})
 \left(1-\frac{1}{4}\frac{m^2_e}{\bar{E}^2_e }\right)  
 \nonumber \\
 &+&\left(\frac{\partial f_{\nu_i}(E'_\nu)}{\partial t}
    \right)_{\nu_i + \bar{\nu}_i\to e^{-}+e^{+}} 
 +\left(\frac{\partial f_{\nu_i}(E'_\nu)}{\partial t}
  \right)_{\nu_i + \bar{\nu}_i \to \mu^{-}+\mu^{+}}
\nonumber \\
 &+&\left(\frac{\partial f_{\nu_e}(E'_\nu)}{\partial t}
    \right)_{\nu_e + \bar{\nu}_\mu\to e^{-}+\mu^{+}}\delta_{i,e}
        +\left(\frac{\partial f_{\nu_\mu}(E'_\nu)}{\partial t}
         \right)_{\nu_\mu + \bar{\nu}_e \to \mu^{-}+e^{+}}\delta_{i,\mu}
        \nonumber \\
 &+&\left(\frac{\partial f_{\nu_e}(E'_\nu)}{\partial t}
    \right)_{\nu_\mu + e^{-}\to \nu_e+\mu^{-}}\delta_{i,e} 
 +\left(\frac{\partial f_{\nu_\mu}(E'_\nu)}{\partial t}
  \right)_{\nu_\mu + e^{-}\to \nu_e+\mu^{-}}\delta_{i,\mu}
 \nonumber \\
 &+&\left(\frac{\partial f_{\nu_i}(E'_\nu)}{\partial t}
    \right)_{\nu_i + \bar{\nu}_i\to \pi^{+} + \pi^{-}} 
 +\left(\frac{\partial f_{\nu_e}(E'_\nu)}{\partial t}
  \right)_{\nu_e + e^{+} \to\pi^{+} + \pi^{0}}\delta_{i,e}
 \nonumber \\
 &+&\frac{1}{2\tau_{X}}n_{X}\delta(E'_\nu -m_{X}/2)\delta_{i,e}
 \nonumber \\
 &+& E'_\nu H \frac{\partial f_{\nu_i}(E'_\nu)}{\partial E'_\nu}
 - 2H f_{\nu_i}(E'_\nu),
\end{eqnarray}
where $H$ is the expansion rate of the universe and $\delta_{i,j}$ is a
Kronecker delta.  The coefficients for neutrino-neutrino scattering are
given by
\begin{eqnarray}
    &&
    a^\nu_{out} = 4,~~~~b^\nu_{out} = 0,~~~~c^\nu_{out} = 13,
    \\ &&
    a^\nu_{in,ii}= 6,~~~b^\nu_{in,ii}= 9,~~~c^\nu_{in,ii}= 11,
    \\ &&
    a^\nu_{in,ij}= 1,~~~b^\nu_{in,ij}= 1,~~~c^\nu_{in,ij}= 2,~~~(i\neq j).
\end{eqnarray}
For example, let us derive $a^\nu_{out}$.  Factor 2 comes from
$\nu_i\nu_i\to\nu_i\nu_i$ scattering and another factor 2 comes from
$\nu_i\nu_j\to\nu_i\nu_j$ scattering.  Consequently, the coefficient
$a^\nu_{out}$ amount to 4.  Other coefficients can be derived in the
same manner.  The coefficients for neutrino-electron scattering are
given by
\begin{eqnarray}
    &&
    a^{e}_{out,e} = (C_V+C_A+2)^2+(C_V-C_A)^2,b^{e}_{out,e} = 0
    ,~~c^{e}_{out,e} = a^{e}_{out,e},
    \\ &&
    a^{e}_{out,j} = (C_V+C_A)^2+(C_V-C_A)^2,~~~~~b^{e}_{out,j} = 0
    ,~~c^{e}_{out,j} = a^{e}_{out,j},
    \\ &&
    a^{e}_{in,i}= a^{e}_{out,i} ,~~~b^{e}_{in,i}= b^{e}_{out,i} 
    ,~~~c^{e}_{in,i}=c^{e}_{out,i},
\end{eqnarray}
where index $i$ represents $e$, $\mu$ and $\tau$ and $j$ represents
$\mu$ and $\tau$.

\section{Inverse Compton}
\label{app:inv}

In this appendix, we write down Boltzmann equations for inverse Compton
process which determine the high energy photon spectrum.  The electron
energies before scattering and after scattering are given by $E_e$ and
$E'_e$.  $E_\gamma$ is used for the energy of scattered  photon and
$\epsilon_\gamma$ for background photon.

The diffuse extragalactic $\gamma$ ray flux has been observed by the
COMPTEL and EGRET measurements.  COMPTEL and EGRET observed $\gamma$ ray
energy ranges from $0.8{\rm MeV}$ to $100{\rm GeV}$.  Electrons and
positrons which scatter up background photons to this energy range
should be highly relativistic. Thus, the number of collisions per unit
time per photon energy through inverse Compton process is
written as~\cite{Jones:1968js},
\begin{eqnarray}
  \frac{d^2N}{dtdE_\gamma}(\epsilon_\gamma,E_\gamma,E_e) 
   &= &  8\pi r_e^2\frac{1}{\Gamma E_e}
   \nonumber \\
   &\times & \left(2q\ln q+(1+2q)(1-q)
              +\frac{1}{2}\frac{(\Gamma q)^2}{1+\Gamma q}(1-q)\right),
   \label{collision_number}
\end{eqnarray}
where $r_e$ is classical electron radius, $\Gamma = 4\epsilon_\gamma
E_e/m_e^2$ and $q = E_\gamma/\Gamma(E_e-E_\gamma)$.  The maximum photon
energy is given by $E_e\Gamma/(1+\Gamma)$.

The Boltzmann equations for inverse Compton process is given by
\begin{eqnarray}
  \frac{\partial f_\gamma(E_\gamma)}{\partial t} &=& 
   \int^\infty_{\frac{E_\gamma}{2}
   \left(1+\sqrt{1+4/\Gamma}\right)} dE_e f_e(E_e)\int^\infty_0
   d\epsilon_\gamma 
   f_b(\epsilon_\gamma)
   \frac{d^2N}{dtdE_\gamma}(\epsilon_\gamma,E_\gamma,E_e) 
   \\
  \frac{\partial f_e(E'_e)}{\partial t} &=& \int^{E_{max}}_{E'_e} 
   dE_e f_e(E_e)\int^\infty_0 d\epsilon_\gamma 
   f_b(\epsilon_\gamma)\frac{d^2N}{dtdE_\gamma}
   (\epsilon_\gamma,E_\gamma=E_e+\epsilon_\gamma-E'_e,E_e) 
   \nonumber
   \\ 
  &-& f_e(E'_e)\int^\infty_0 d\epsilon_\gamma f_b(\epsilon_\gamma)
   \frac{dN}{dt}(\epsilon_\gamma,E'_e)
\end{eqnarray}
where $E_{max} = {m_e}^2(E'_e-\epsilon_\gamma)
/(4{\epsilon_\gamma}^2+{m_e}^2-4\epsilon_\gamma E'_e)$ when the
denominator is positive and $E_{max} = \infty$ when negative.  The
distribution function of the background photons at temperature
$T_\gamma$ is represented by $f_b(\epsilon_\gamma)$,
\begin{eqnarray}
   f_b(\epsilon_\gamma) = \frac{{\epsilon_\gamma}^2}{\pi^2}
    \frac{1}{\exp(\epsilon_\gamma/T_\gamma)-1}.
\end{eqnarray}  
To check the validity of these Boltzmann equations, we show the electron
number conservation.
\begin{eqnarray*}
   \fl \int^\infty_{m_e} dE'_e \frac{\partial f_e(E'_e)}{\partial t} 
    = \int^\infty_{m_e} dE_e f_e(E_e)
    \int^{E_e}_{E_e/(1+\Gamma)+\epsilon_\gamma} dE'_e
    \int^\infty_0 d\epsilon_\gamma f_b(\epsilon_\gamma)
    \frac{d^2N}{dtdE_\gamma}
    (\epsilon_\gamma,E_e+\epsilon_\gamma-E'_e,E_e) 
    \\
   -\int^{\infty}_{m_e}dE'_e f_e(E'_e)\int^\infty_0 
    d\epsilon_\gamma f_b(\epsilon_\gamma)
    \frac{dN}{dt}(\epsilon_\gamma,E'_e) 
    \\
   = \int^\infty_{m_e} dE_e f_e(E_e)
    \int^{E_e\Gamma/(1+\Gamma)}_{\epsilon_\gamma}dE_\gamma
    \int^\infty_0 d\epsilon_\gamma f_b(\epsilon_\gamma)
    \frac{d^2N}{dtdE_\gamma}
    (\epsilon_\gamma,E_\gamma,E_e) \nonumber 
    \\
   -\int^{\infty}_{m_e}dE'_e f_e(E'_e)\int^\infty_0 
    d\epsilon_\gamma f_b(\epsilon_\gamma)
    \frac{dN}{dt}(\epsilon_\gamma,E'_e) 
    \\
   = \int^{\infty}_{m_e}dE_e f_e(E_e)\int^\infty_0 
    d\epsilon_\gamma f_b(\epsilon_\gamma)
    \frac{dN}{dt}(\epsilon_\gamma,E_e) 
    \\
   - \int^{\infty}_{m_e}dE'_e f_e(E'_e)\int^\infty_0 
    d\epsilon_\gamma f_b(\epsilon_\gamma)
    \frac{dN}{dt}(\epsilon_\gamma,E'_e) 
    \\
   =   0
\end{eqnarray*} 
The energy conservation is also easily shown.
\begin{eqnarray*}
   \fl \int^\infty_{\epsilon_\gamma} 
    dE_\gamma(E_\gamma-\epsilon_\gamma)
    \frac{\partial f_\gamma(E_\gamma)}{\partial t} 
    =  \int^{\infty}_{m_e}dE_e f_e(E_e)
    \int^\infty_0 d\epsilon_\gamma f_b(\epsilon_\gamma)  
    \\
   \times
    \int^{E_e\Gamma/(1+\Gamma)}_{\epsilon_\gamma}
    dE_\gamma (E_\gamma-\epsilon_\gamma)
    \frac{d^2N}{dtdE_\gamma}(\epsilon_\gamma,E_\gamma,E_e)
    \\
   \fl \int^\infty_{m_e} 
    dE'_e E'_e\frac{\partial f_e(E'_e)}{\partial t} = 
    \int^{\infty}_{m_e}dE_e f_e(E_e)
    \int^\infty_0 d\epsilon_\gamma f_b(\epsilon_\gamma)
    \\
   \times 
    \int^{E_e\Gamma/(1+\Gamma)}_{\epsilon_\gamma}dE_\gamma 
    (E_e+\epsilon_\gamma-E_\gamma)
    \frac{d^2N}{dtdE_\gamma}(\epsilon_\gamma,E_\gamma,E_e) 
    \\
   - \int^{\infty}_{m_e}dE'_e E'_ef_e(E'_e)\int^\infty_0 
    d\epsilon_\gamma f_b(\epsilon_\gamma)
    \frac{dN}{dt}(\epsilon_\gamma,E'_e) 
    \\
   = -\int^\infty_{\epsilon_\gamma} 
    dE_\gamma(E_\gamma-\epsilon_\gamma)
    \frac{\partial f_\gamma(E_\gamma)}{\partial t} 
\end{eqnarray*}  


\begin{figure}[thbp]
  \begin{center}
    \includegraphics[width=0.8\linewidth]{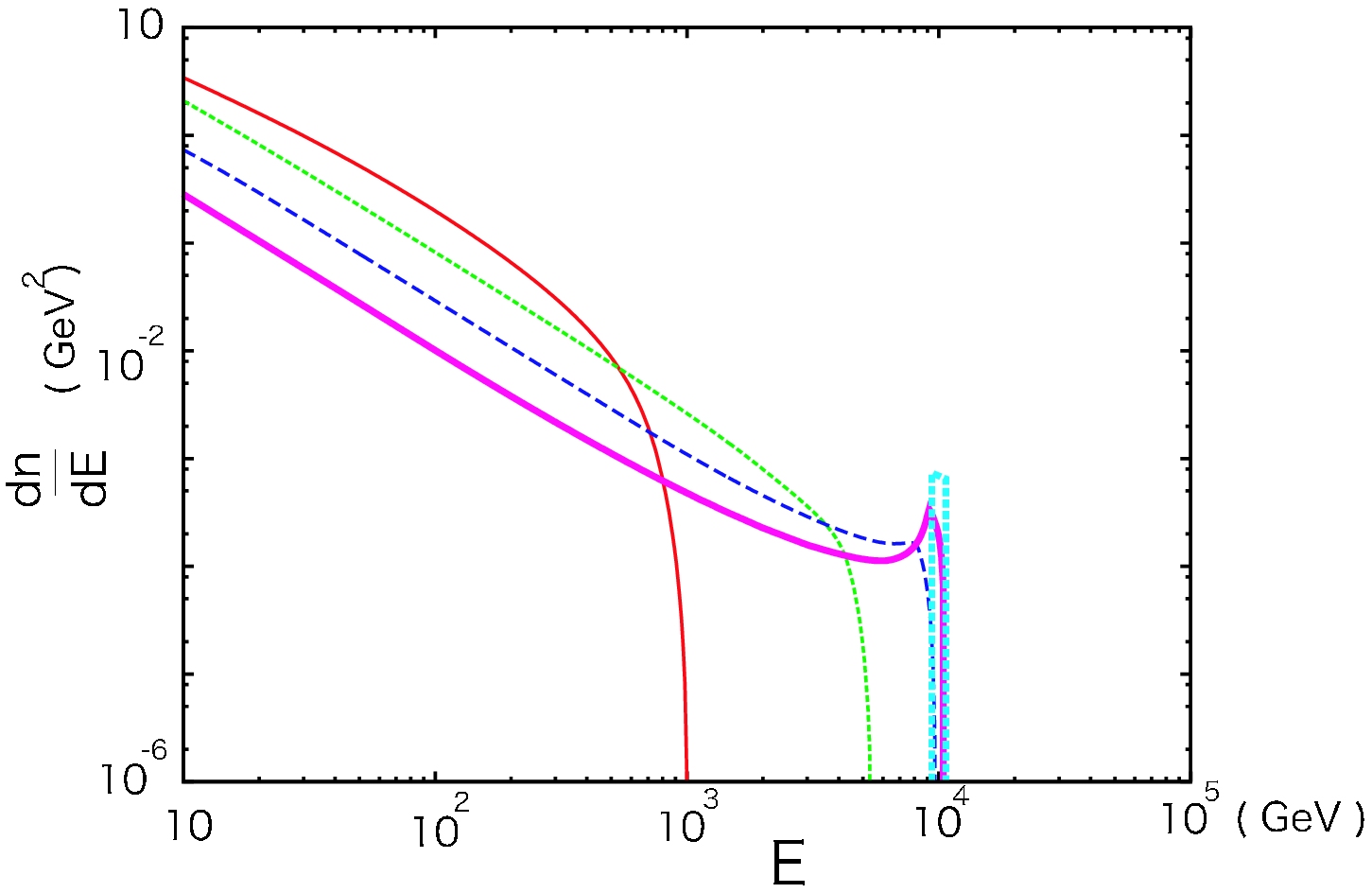}
     \caption{Photon spectra which are produced through inverse Compton
     process when high energy electrons are injected with $E_e =
     10^4~{\rm GeV}$ and $n_e = 1~{\rm GeV}^3$.  From upper to lower the
     lines represent the photon spectra when $1+z=1$ (thin solid line),
     $10$ (thin dotted line), $100$ (thin dashed line) and $1000$ (thick
     solid line).
     Thick dotted line represents the initial electron
     spectrum.
     For simplicity, we have assumed here that background photon spectrum
     is monochromatic with $\epsilon_\gamma=2.7T_\gamma$}  \label{fig:inverse_compton}
 \end{center}
\end{figure}


In Fig.~\ref{fig:inverse_compton}, we plot the photon spectra which are
produced through inverse Compton process when high energy electrons are
injected. 
The photon spectra
have quite a different form for different value of $\Gamma$.  In the
Thomson limit corresponding to $\Gamma\ll1$, the first two terms of the
right hand side of Eq.~(\ref{collision_number}) is dominant.  In this
case, many low-energy photons are produced.  In the extreme
Klein-Nishina limit corresponding to $\Gamma\gg1$, the last term of the
right hand side of Eq.~(\ref{collision_number}) is dominant at larger
$q$ and photon spectra have peaks near the high-energy
end~\cite{Blumenthal:1970gc}.  Our calculation is different from the
steady-state method~\cite{Blumenthal:1970gc,Blumenthal:1971fh}.
However, once electron spectrum is in a steady-state, our result is in
good agreement with the result
in~\cite{Blumenthal:1970gc,Blumenthal:1971fh}.

\end{document}